\documentclass[twocolumn,english,showpacs,preprintnumbers,prb]{revtex4-1}
\usepackage[T1]{fontenc}
\usepackage[latin9]{inputenc}
\usepackage{amsmath}
\usepackage{amssymb}
\usepackage{graphicx}
\usepackage{esint}
\usepackage{color}

\makeatletter

\@ifundefined{textcolor}{}{%
 \definecolor{BLACK}{gray}{0}
 \definecolor{WHITE}{gray}{1}
 \definecolor{RED}{rgb}{1,0,0}
 \definecolor{GREEN}{rgb}{0,1,0}
 \definecolor{BLUE}{rgb}{0,0,1}
 \definecolor{CYAN}{cmyk}{1,0,0,0}
 \definecolor{MAGENTA}{cmyk}{0,1,0,0}
 \definecolor{YELLOW}{cmyk}{0,0,1,0}
 }

\usepackage{epsfig}\usepackage{dcolumn}\usepackage{bm}

\usepackage{babel}

\makeatother

\usepackage{babel}
\begin{document}

\title{Dimensionality effects in the LDOS  of ferromagnetic
hosts probed via STM: spin-polarized quantum beats and spin filtering}

\author{A. C. Seridonio$^{1,2}$, S. C. Leandro$^{1}$, L. H. Guessi$^{1}$,
E. C. Siqueira$^{1}$, F. M. Souza$^{3}$, E. Vernek$^{3}$, M. S.
Figueira$^{4}$, and J. C. Egues$^{5}$}

\affiliation{$^{1}$Departamento de F\'{i}sica e Qu\'{i}mica, Universidade Estadual Paulista, 15385-000, Ilha Solteira,
SP, Brazil\\
$^{2}$Instituto de Geoci\^encias e Ci\^encias Exatas - IGCE, Universidade Estadual Paulista, Departamento
de F\'{i}sica, 13506-970, Rio Claro, SP, Brazil \\
$^{3}$Instituto de F\'{i}sica, Universidade Federal de Uberl\^andia,
38400-902, Uberl\^andia, MG, Brazil.\\
$^{4}$Instituto de F\'{i}sica, Universidade Federal Fluminense,
24210-340 Niter\'oi, RJ, Brazil \\
$^{5}$Instituto de F\'{i}sica de S\^ao Carlos, Universidade de S\~ao
Paulo, 13560-970, S\~ao Carlos, SP, Brazil. }
\begin{abstract}
We theoretically investigate the local density of states (LDOS) probed by a STM tip of ferromagnetic
metals hosting a single adatom and a subsurface impurity. We model the system via the two-impurity Anderson
Hamiltonian. By using the equation of motion with the relevant Green functions, we derive analytical
expressions for the LDOS of two host types: a surface and a quantum
wire. The LDOS reveals Friedel-like oscillations and Fano interference as a function of the STM tip position. These oscillations strongly
depend on the host dimension. Interestingly, we find that the spin-dependent Fermi wave numbers of the hosts give rise to spin-polarized \textit{quantum
beats} in the LDOS. While the LDOS for the metallic surface shows a damped beating pattern, it exhibits an opposite behavior in the quantum wire. Due to this absence of damping, the wire operates as a
spatially resolved spin filter with a high efficiency.
\end{abstract}


\maketitle

\section{Introduction}

\label{sec1}

The local density of states (LDOS) of electronic systems with impurities
can exhibit Fano line shapes, due to the quantum interference between
different electron paths. Such interference arises from the itinerant
electrons that travel through the host and tunnel into the impurity
sites. \cite{Fano1,Fano2} For a single magnetic adatom in the Kondo
regime \cite{Hewson} probed by a scanning tunneling microscope (STM)
tip, interesting features manifest when one has a spin-polarized electron bath present.
Here we mention the splitting of the Kondo peak in the differential conductance due to the itinerant magnetism of the host.\cite{Seridonio}Such
hallmark has already been found experimentally in an Fe island with
a Co adatom.\cite{Kawahara} Additionally, the STM system can also operate as a Fano-Kondo spin-filter due to a spin-polarized tip and a nonmagnetic host.\cite{SPSTM1,SPSTM2}In the absence of a ferromagnetic host, the Fano-Kondo profile becomes doubly degenerate. \cite{STM1,STM2,STM3,STM4,STM5,STM6,STM7,AHCNeto,STM8,STM9,STM10,STM11,STM12,STM13,STM14,STM15}
Away from the Kondo regime, a spin diode emerges.\cite{Poliana}

In the condensed matter literature on scanning microscopy, there is a profusion of
work discussing spin-dependent phenomena  employing
ferromagnetic leads coupled to quantum dots or adatoms in the Kondo regime. \cite{SPSTM5,FM1,Yunong,FM2,FM3,FM4,FM5,FM6,FM7,FM8,FM88,FM9,FM10,FM11,FM12,new1,new2,new3,new4,new5,SPSTM1,SPSTM2,Seridonio}  Here we mention those with metallic samples and buried impurities, in which the anisotropy of the Fermi surface plays an important role in electron tunneling. \cite{STM16,STM17,STM18_FS,STM19_FS,STM20_FS,STM21_FS}
According to the experiment of Pr\"user \textit{et al.} \cite{STM16},
such anisotropy allows atoms of Fe and Co beneath the Cu(100) surface to scatter electrons
in preferential directions of the material due to an effect called  ``electron focusing''. In this scenario, the STM becomes a new tool for the detection of the
Fermi surface signatures in the real lattice of a metal. In contrast, much less attention has been devoted to spin-polarized systems away from the Kondo regime and with two impurities.

Thus we present in this work a theoretical description of the systems sketched in Fig. \ref{fig:Pic1}. We show that interesting phenomena such as the spin-polarized quantum beats in the LDOS and the spin-filtering effect arise. To this end, we consider two distinct geometries consistent with recent experiments: a metallic surface and a quantum wire. The 2D case emulates the Fe island in Ref.{[}\onlinecite{Kawahara}{]}. The quantum wire on the other hand, mimics the ``electron focusing'' effect investigated in  Ref.{[}\onlinecite{STM16}{]. Interestingly, we note that the pioneering quantum wire treatment for ``electron focusing'' in a side-coupled geometry can be found in Ref.{[}\onlinecite{STM20_FS}{]}. In this treatment the noninteracting single impurity Anderson model\cite{SIAM} was solved in one dimension by considering the impurity above the wire. We should also point out that the full ab-initio calculation that yields to ``electron focusing'' in Ref.{[}\onlinecite{STM16}{]} can be qualitatively recovered by the simple quantum wire model adopted in Ref.{[}\onlinecite{STM20_FS}{]}.

Here we extend this one-dimensional treatment of the Anderson Hamiltonian by including a spin-dependent DOS for the wire, a second lateral impurity right beneath it and Coulomb interaction in both impurities. We perform our study in the framework of the two-impurity Anderson model by employing the equation of motion approach to calculate the LDOS of the system. The Hubbard I approximation\cite{book2} is used by assuming for the sake of simplicity infinite Coulomb energies at the impurities. We show that the LDOS can be written in terms of the Fano factor, the Friedel-like function for charge oscillations and the spin-dependent Fermi wave numbers of the host. Such quantities lead to spin-polarized quantum beats in the LDOS.
We also show that this effect is strongly correlated to the host dimensionality. Thus the quantum beats in the LDOS of the metallic surface present a long-range
damped behavior in contrast to the undamped one found in the quantum wire system.
Such distinct features originate from the specific forms assumed by the Fano factor and Friedel function, which depend on the dimensionality of the host. Therefore the metallic surface and the quantum wire become spatially resolved spin filters, where the latter displays a higher efficiency due to the undamped LDOS.

\begin{figure}[tbh]
 \includegraphics[clip,width=0.48\textwidth]{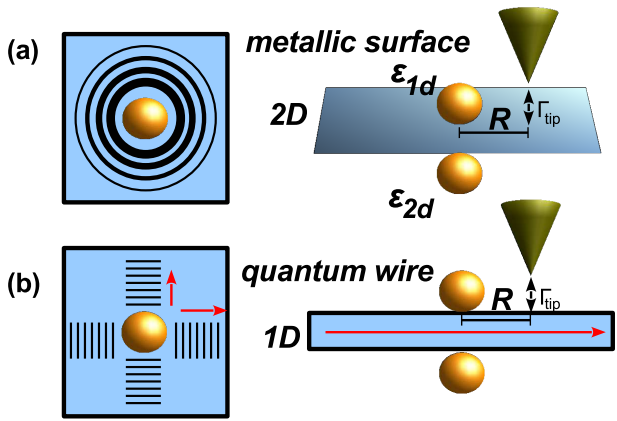} \caption{\label{fig:Pic1}(Color online) Side-coupled geometry with two impurities in presence
of a STM tip. $\Gamma_{tip}$ is the tip-host coupling. a) left panel: 2D evanescent waves appear in the LDOS of a metallic surface. The system is treated as a two-dimensional electron gas showed in the right panel. b) left panel: the confinement of 1D waves in specific directions (perpendicular
wave fronts) is due to the ``electron focusing'' effect (see Ref.{[}\onlinecite{STM16}{]}). Each direction is modeled by a quantum wire as illustrated
in the right panel.}
\end{figure}

This paper is organized as follows. In Sec. \ref{sec2}, we show the theoretical model of the ferromagnetic hosts with the impurities in the side-coupled geometry as sketched in Fig. \ref{fig:Pic1} and derive the LDOS formula for both systems, the metallic surface and the quantum wire. The decoupling scheme
Hubbard I \cite{book2} for the Green functions is presented in Sec. \ref{sec3}. In Sec. \ref{sec4}, we discuss the results for
the quantum beats in the LDOS and the spin-filtering. The conclusions appear in Sec. \ref{sec5}.

\section{Theoretical Model}

\label{sec2}

\subsection{Hamiltonian}

\label{sub:sec2A}

In order to probe the LDOS of the ferromagnetic hosts, we represent a STM tip
weakly connected to hosts hybridized to a pair of side-coupled impurities
as outlined in Fig. \ref{fig:Pic1}. The systems we investigate are
described according to the two-impurity Anderson model given by the Hamiltonian\cite{Hewson}

\begin{align}
\mathcal{H} & =\sum_{\vec{k}\sigma}\varepsilon_{\vec{k}\sigma}c_{\vec{k}\sigma}^{\dagger}c_{\vec{k}\sigma}+\sum_{j\sigma}\varepsilon_{jd\sigma}d_{j\sigma}^{\dagger}d_{j\sigma}+\sum_{j}U_{j}d_{j\uparrow}^{\dagger}d_{j\uparrow}d_{j\downarrow}^{\dagger}d_{j\downarrow}\nonumber \\
 & +\sum_{j\vec{k}\sigma}\left[\frac{V_{j\vec{k}\sigma}}{\sqrt{\mathcal{N}_{\sigma}}}\phi_{\vec{k}\sigma}\left(\vec{R}_{j}\right)c_{\vec{k}\sigma}^{\dagger}d_{j\sigma}+\frac{V_{j\vec{k}\sigma}^{*}}{\sqrt{\mathcal{N}_{\sigma}}}\phi_{\vec{k}\sigma}^{*}\left(\vec{R}_{j}\right)d_{j\sigma}^{\dagger}c_{\vec{k}\sigma}\right],\label{eq:TIAM}
\end{align}
 The spin-polarized electron gas forming the hosts is described by the operator
$c_{\vec{k}\sigma}^{\dagger}$ ($c_{\vec{k}\sigma}$) for the creation
(annihilation) of an electron in a quantum state labeled by the wave
vector $\vec{k}$, spin $\sigma$ and energy $\varepsilon_{\vec{k}\sigma}.$
For the impurities, $d_{j\sigma}^{\dagger}$ ($d_{j\sigma}$) creates
(annihilates) an electron with spin $\sigma$ in the state $\varepsilon_{jd\sigma}$,
with $j=1,2.$ The third term of Eq.~(\ref{eq:TIAM}) accounts for the on-site
Coulomb interaction $U_{j}$ at the $j$th impurity placed at position
$\vec{R}_{j}$. In our calculations, we assume $U_{1}=U_{2}\rightarrow\infty$
(single occupancy of the impurities). Finally, the last two terms mix the host continuum of states and
the levels $\varepsilon_{jd\sigma}.$ This hybridization occurs at
the impurity sites $\vec{R}_{j}$ via the host-impurity couplings $V_{j\vec{k}\sigma}$
and the plane waves $\phi_{\vec{k}\sigma}\left(\vec{R}_{j}\right)=e^{i\vec{k}.\vec{R_{j}}}$.
$\mathcal{N}_{\sigma}$ is the number of conduction states for a given
spin $\sigma$. The ferromagnetic hosts are considered spin-polarized electron baths, characterized by the polarization
\begin{equation}
P=\frac{\rho_{FM\uparrow}-\rho_{FM\downarrow}}{\rho_{FM\uparrow}+\rho_{FM\downarrow}},\label{eq:SP}
\end{equation}
 in which
\begin{equation}
\rho_{FM\sigma}=\frac{1}{2D_{\sigma}}=\rho_{0}(1+\sigma P)\label{eq:rho_spin}
\end{equation}
 is the density of states of the hosts in a Stoner-like framework\cite{Stoner,BJP} expressed in terms of  the spin-dependent half-width $D_{\sigma}$ and the density $\rho_{0}$
for the case $P=0$.

\subsection{LDOS for the spin-polarized systems}

\label{sub:sec2B} To obtain the host LDOS we introduce the retarded
Green function in the time coordinate,
\begin{align}
\mathcal{G}_{\sigma}\left(t,\vec{R}\right) & =-\frac{i}{\hbar}\theta\left(t\right)\mathcal{Z}_{FM}^{-1}\sum_{n}e^{-\beta E_{n}}\nonumber \\
 & \times\left\langle n\right|\left[\tilde{\Psi}_{\sigma}\left(\vec{R},t\right),\tilde{\Psi}_{\sigma}^{\dagger}\left(\vec{R},0\right)\right]_{+}\left|n\right\rangle ,\label{eq:PSI_R}
\end{align}
 where
\begin{equation}
\tilde{\Psi}_{\sigma}\left(\vec{R}\right)=\frac{1}{\sqrt{\mathcal{N}_{\sigma}}}\sum_{\vec{k}}\phi_{\vec{k}\sigma}\left(\vec{R}\right)c_{\vec{k}\sigma}\label{eq:PSI_R-1-1}
\end{equation}
 is the fermionic operator describing the quantum state of the host
site placed below the STM tip, $\hbar$ is the Planck constant divided
by $2\pi$, $\theta\left(t\right)$ the step function at the instant
$t$, $\beta=1/k_{B}T$ with $k_{B}$ as the Boltzmann constant
and $T$ the system temperature, $\mathcal{Z}_{FM}$ and $\left|n\right\rangle $
are the partition function and a many-body eigenstate of the system Hamiltonian
{[}Eq. (\ref{eq:TIAM}){]}, respectively, and $[\cdots,\cdots]_{+}$
is the anticommutator for Eq. (\ref{eq:PSI_R-1-1}) evaluated at distinct
times. From Eq.~(\ref{eq:PSI_R}) the spin-dependent LDOS at a site
$\vec{R}$ of the host {[}see Fig. \ref{fig:Pic1}{]} can be obtained
as
\begin{equation}
\rho_{LDOS}^{\sigma}\left(\varepsilon,R\right)=-\frac{1}{\pi}{\tt Im}\left\{ \tilde{\mathcal{G}}_{\sigma}\left(\varepsilon^{+},\vec{R}\right)\right\} ,\label{eq:FM_LDOS}
\end{equation}
 where $\tilde{\mathcal{G}}_{\sigma}(\varepsilon^{+},\vec{R})$ is
the time Fourier transform of $\mathcal{G}_{\sigma}(t,\vec{R})$.
Here, $\varepsilon^{+}=\varepsilon+i\eta$ and $\eta\rightarrow0^{+}$.
In what follows we first develop a general formalism for impurities
localized at arbitrary positions $\vec{R}_{j}$ and $\vec{R}_{l}$;
later on we take the limit $\vec{R}_{j}=\vec{R}_{l}=\vec{0}$, in
order to treat the side-coupled geometry of this work.

To obtain an analytical expression for the LDOS we apply the equation-of-motion approach to Eq. (\ref{eq:PSI_R}). Thus we substitute
Eq. (\ref{eq:PSI_R-1-1}) in Eq. (\ref{eq:PSI_R}) and begin the procedure
with
\begin{equation}
\mathcal{G}{}_{\sigma}\left(t,\vec{R}\right)=\frac{1}{\mathcal{N}_{\sigma}}\sum_{\vec{k}\vec{q}}\phi_{\vec{k}\sigma}\left(\vec{R}\right)\phi_{\vec{q}\sigma}^{*}\left(\vec{R}\right)\mathcal{G}_{c_{\vec{k}}c_{\vec{q}}}^{\sigma}\left(t\right)\label{eq:GF_1}
\end{equation}
 expressed in terms of
\begin{align}
\mathcal{G}_{c_{\vec{k}}c_{\vec{q}}}^{\sigma}\left(t\right) & =-\frac{i}{\hbar}\theta\left(t\right)\mathcal{Z}_{FM}^{-1}\sum_{n}e^{-\beta E_{n}}\nonumber \\
 & \times\left\langle n\right|\left[c_{\vec{k}\sigma}\left(t\right),c_{\vec{q}\sigma}^{\dagger}\left(0\right)\right]_{+}\left|n\right\rangle .\label{eq:GF_2}
\end{align}
 Performing $\frac{\partial}{\partial t}$ on Eq. (\ref{eq:GF_2})
we find
\begin{eqnarray}
\frac{\partial}{\partial t}\mathcal{G}_{c_{\vec{k}}c_{\vec{q}}}^{\sigma}\left(t\right) & = & -\frac{i}{\hbar}\delta\left(t\right)\mathcal{Z}_{FM}^{-1}\sum_{n}e^{-\beta E_{n}}\nonumber \\
 & \times & \left\langle n\right|\left[c_{\vec{k}\sigma}\left(t\right),c_{\vec{q}\sigma}^{\dagger}\left(0\right)\right]_{+}\left|n\right\rangle \nonumber \\
 & + & \left(-\frac{i}{\hbar}\right)\varepsilon_{\vec{k}\sigma}\mathcal{G}_{c_{\vec{k}}c_{\vec{q}}}^{\sigma}\left(t\right)+\left(-\frac{i}{\hbar}\right)\nonumber \\
 & \times & \frac{1}{\sqrt{\mathcal{\mathcal{N}_{\sigma}}}}\sum_{j}V_{j\vec{k}\sigma}\phi_{\vec{k}\sigma}^{*}\left(\vec{R}_{j}\right)\mathcal{G}_{d_{j}c_{\vec{q}}}^{\sigma}\left(t\right),\label{eq:GF_3}
\end{eqnarray}
 where we have used
\begin{align}
i\hbar\frac{\partial}{\partial t}c_{\vec{k}\sigma}\left(t\right) & =\left[c_{\vec{k}\sigma},\mathcal{H}\right]=\varepsilon_{\vec{k}\sigma}c_{\vec{k}\sigma}\left(t\right)\nonumber \\
 & +\frac{1}{\sqrt{\mathcal{\mathcal{N}_{\sigma}}}}\sum_{j}V_{j\vec{k}\sigma}\phi_{\vec{k}\sigma}^{*}\left(\vec{R}_{j}\right)d_{j\sigma}\left(t\right).\label{eq:HB_I}
\end{align}

In the energy coordinate, we solve Eq. (\ref{eq:GF_3}) for $\tilde{\mathcal{G}}_{c_{\vec{k}}c_{\vec{q}}}^{\sigma}\left(\varepsilon^{+}\right)$
and obtain

\begin{align}
\tilde{\mathcal{G}}_{c_{\vec{k}}c_{\vec{q}}}^{\sigma}\left(\varepsilon^{+}\right) & =\frac{\delta_{\vec{k}\vec{q}}}{\varepsilon^{+}-\varepsilon_{\vec{k}\sigma}}+\frac{1}{\sqrt{\mathcal{N}_{\sigma}}}\sum_{j}\frac{V_{\vec{jk}\sigma}\phi_{\vec{k}\sigma}^{*}\left(\vec{R}_{j}\right)}{\varepsilon^{+}-\varepsilon_{\vec{k}\sigma}}\nonumber \\
 & \times\tilde{\mathcal{G}}_{d_{j}c_{\vec{q}}}^{\sigma}\left(\varepsilon^{+}\right).\label{eq:GF_4}
\end{align}

Notice that we need to find the mixed Green function, $\tilde{\mathcal{G}}_{d_{j}c_{\vec{q}}}^{\sigma}\left(\varepsilon^{+}\right).$
To this end, we define the advanced Green function

\begin{align}
\mathcal{F}_{d_{j}c_{\vec{q}}}^{\sigma}\left(t\right) & =\frac{i}{\hbar}\theta\left(-t\right)\mathcal{Z}_{FM}^{-1}\sum_{n}e^{-\beta E_{n}}\nonumber \\
 & \times\left\langle n\right|\left[d_{j\sigma}^{\dagger}\left(0\right),c_{\vec{q}\sigma}\left(t\right)\right]_{+}\left|n\right\rangle ,\label{eq:GF_5}
\end{align}
 which results in
\begin{align}
\frac{\partial}{\partial t}\mathcal{F}_{d_{j}c_{\vec{q}}}^{\sigma}\left(t\right) & =-\frac{i}{\hbar}\delta\left(t\right)\mathcal{Z}_{FM}^{-1}\sum_{n}e^{-\beta E_{n}}\nonumber \\
 & \times\left\langle n\right|\left[d_{j\sigma}^{\dagger}\left(0\right),c_{\vec{q}\sigma}\left(t\right)\right]_{+}\left|n\right\rangle -\frac{i}{\hbar}\varepsilon_{\vec{q}\sigma}\mathcal{F}_{d_{j}c_{\vec{q}}}^{\sigma}\left(t\right)\nonumber \\
 & +\left(-\frac{i}{\hbar}\right)\frac{1}{\sqrt{\mathcal{\mathcal{N}_{\sigma}}}}\sum_{l}V_{l\vec{q}\sigma}\phi_{\vec{q}\sigma}^{*}\left(\vec{R}_{l}\right)\mathcal{F}_{d_{j}d_{l}}^{\sigma}\left(t\right),\label{eq:GF_6}
\end{align}
 where we used once again Eq. (\ref{eq:HB_I}), interchanging $\vec{k}\leftrightarrow\vec{q}$.
Thus the Fourier transform of Eq.~(\ref{eq:GF_6}) becomes
\begin{align}
\varepsilon^{-}\tilde{\mathcal{F}}_{d_{j}c_{\vec{q}}}^{\sigma}\left(\varepsilon^{-}\right) & =\varepsilon_{\vec{q}\sigma}\tilde{\mathcal{F}}_{d_{j}c_{\vec{q}}}^{\sigma}\left(\varepsilon^{-}\right)+\frac{1}{\sqrt{\mathcal{\mathcal{N}_{\sigma}}}}\sum_{l}V_{l\vec{q}\sigma}\phi_{\vec{q}\sigma}^{*}\left(\vec{R}_{l}\right)\nonumber \\
 & \times\tilde{\mathcal{F}}_{d_{j}d_{l}}^{\sigma}\left(\varepsilon^{-}\right),\label{eq:GF_7}
\end{align}
 with $\varepsilon^{-}=\varepsilon-i\eta$. Applying the property
\begin{equation}
\tilde{\mathcal{G}}_{d_{j}c_{\vec{q}}}^{\sigma}\left(\varepsilon^{+}\right)=\left\{ \tilde{\mathcal{F}}_{d_{j}c_{\vec{q}}}^{\sigma}\left(\varepsilon^{-}\right)\right\} ^{\dagger}\label{eq:property}
\end{equation}
 to Eq. (\ref{eq:GF_7}), we show that
\begin{align}
\varepsilon^{+}\tilde{\mathcal{G}}_{d_{j}c_{\vec{q}}}^{\sigma}\left(\varepsilon^{+}\right) & =\varepsilon_{\vec{q}\sigma}\tilde{\mathcal{G}}_{d_{j}c_{\vec{q}}}^{\sigma}\left(\varepsilon^{+}\right)+\frac{1}{\sqrt{\mathcal{\mathcal{N}_{\sigma}}}}\sum_{l}V_{l\vec{q}\sigma}^{*}\phi_{\vec{q}\sigma}\left(\vec{R}_{l}\right)\nonumber \\
 & \times\tilde{\mathcal{G}}_{d_{j}d_{l}}^{\sigma}\left(\varepsilon^{+}\right)\label{eq:GF_8}
\end{align}
 and
\begin{equation}
\tilde{\mathcal{G}}_{d_{j}c_{\vec{q}}}^{\sigma}\left(\varepsilon^{+}\right)=\frac{1}{\sqrt{\mathcal{\mathcal{N}_{\sigma}}}}\sum_{l}\frac{V_{l\vec{q}\sigma}^{*}\phi_{\vec{q}\sigma}\left(\vec{R}_{l}\right)}{\varepsilon^{+}-\varepsilon_{\vec{q}\sigma}}\tilde{\mathcal{G}}_{d_{j}d_{l}}^{\sigma}\left(\varepsilon^{+}\right).\label{eq:GF_9}
\end{equation}

Now we substitute Eq. (\ref{eq:GF_9}) into Eq. (\ref{eq:GF_4}) and
the latter into Eq. (\ref{eq:GF_1}) in the energy coordinate to obtain
\begin{eqnarray}
\tilde{\mathcal{G}}_{\sigma}\left(\varepsilon^{+},R\right) & = & \frac{1}{\mathcal{\mathcal{N}_{\sigma}}}\sum_{\vec{k}}\frac{\left|\phi_{\vec{k}\sigma}\left(\vec{R}\right)\right|^{2}}{\varepsilon^{+}-\varepsilon_{\vec{k}\sigma}}\nonumber \\
 & + & \left(\pi\rho_{0}\right)^{2}\sum_{j}\left(q_{j\sigma}-iA_{j\sigma}\right)\left(q_{j}-iA_{j\sigma}\right)\nonumber \\
 & \times & \tilde{\mathcal{G}}_{d_{j}d_{j}}^{\sigma}\left(\varepsilon\right)\nonumber \\
 & + & \left(\pi\rho_{0}\right)^{2}\sum_{j\neq l}\left(q_{j\sigma}-iA_{j\sigma}\right)\left(q_{l\sigma}-iA_{l\sigma}\right)\nonumber \\
 & \times & \tilde{\mathcal{G}}_{d_{j}d_{l}}^{\sigma}\left(\varepsilon\right).\label{eq:GFN}
\end{eqnarray}
 It is worth mentioning that the imaginary part of the first term
of Eq. (\ref{eq:GFN}) gives the background DOS of the host {[}Eq.
(\ref{eq:rho_spin}){]} and the others describe impurity contributions,
with

\begin{equation}
q_{j\sigma}=\frac{1}{\pi\rho_{0}\mathcal{\mathcal{\mathcal{N}_{\sigma}}}}\sum_{\vec{k}}\frac{V_{j\vec{k}\sigma}\phi_{\vec{k}\sigma}^{*}\left(\vec{R}_{j}\right)\phi_{\vec{k}\sigma}\left(\vec{R}\right)}{\varepsilon-\varepsilon_{\vec{k}\sigma}}\label{eq:Fano_j}
\end{equation}
 being the Fano parameter due to the single coupling $V_{j\vec{k}\sigma}$
between the host and a given impurity. This factor encodes the quantum
interference originated by electrons traveling through the ferromagnetic conduction band that tunnel to the impurity state and return to the
band, and those that do not perform such trajectory. Additionally,
we recognize

\begin{align}
A_{j\sigma} & =\frac{1}{\rho_{0}\mathcal{\mathcal{\mathcal{N}_{\sigma}}}}\sum_{\vec{k}}V_{j\vec{k}\sigma}\phi_{\vec{k}\sigma}^{*}\left(\vec{R}_{j}\right)\phi_{\vec{k}\sigma}\left(\vec{R}\right)\delta\left(\varepsilon-\varepsilon_{\vec{k}\sigma}\right)\nonumber \\
 & =\left|A_{j\sigma}\right|e^{i\alpha_{j\sigma}}\label{eq:Friedel_j}
\end{align}
 as an expression that we call the Friedel function, because it leads
to Friedel-like oscillations in the LDOS with $\alpha_{j\sigma}$
as a spin-dependent phase. At the end, the Green function $\tilde{\mathcal{G}}_{\sigma}\left(\varepsilon^{+},R\right)$
indeed depends on $\tilde{\mathcal{G}}_{d_{j}d_{j}}^{\sigma}\left(\varepsilon\right)$
and the mixed Green function $\tilde{\mathcal{G}}_{d_{j}d_{l}}^{\sigma}\left(\varepsilon\right)$.
Finally, from Eqs.~(\ref{eq:FM_LDOS}) and (\ref{eq:GFN}), the LDOS
of the ferromagnetic systems can be recast as the expression

\begin{eqnarray}
\rho_{LDOS}^{\sigma}\left(\varepsilon,R\right) & = & \rho_{FM\sigma}+\pi\rho_{0}^{2}\sum_{j}\left[\left(\left|A_{j\sigma}\right|^{2}-q_{j\sigma}^{2}\right)\right.\nonumber \\
 & \times & {\tt Im}\left\{ \tilde{\mathcal{G}}_{d_{j}d_{j}}^{\sigma}\left(\varepsilon\right)\right\} +2q_{j\sigma}\left|A_{j\sigma}\right|\nonumber \\
 & \times & \left.\sin\left(\alpha_{j\sigma}+\frac{\pi}{2}\right){\tt Re}\left\{ \tilde{\mathcal{G}}_{d_{j}d_{j}}^{\sigma}\left(\varepsilon\right)\right\} \right]\nonumber \\
 & + & \pi\rho_{0}^{2}\sum_{j\neq l}\Delta\varrho_{jl\sigma},\label{eq:LDOS_p1}
\end{eqnarray}
 where
\begin{eqnarray}
\Delta\varrho_{jl\sigma} & = & -q_{j\sigma}q_{l\sigma}{\tt Im}\left\{ \tilde{\mathcal{G}}_{d_{j}d_{l}}^{\sigma}\left(\varepsilon\right)\right\} +\left[\left|A_{j\sigma}\right|q_{l\sigma}\right.\nonumber \\
 & \times & \cos\left(\alpha_{j\sigma}+\frac{\pi}{2}\right)+\left|A_{j\sigma}\right|\left|A_{l\sigma}\right|\cos\left(\alpha_{j\sigma}-\alpha_{l\sigma}\right)\nonumber \\
 & - & \left.q_{j\sigma}\left|A_{l\sigma}\right|\cos\left(\alpha_{l\sigma}+\frac{\pi}{2}\right)\right]{\tt Im}\left\{ \tilde{\mathcal{G}}_{d_{j}d_{l}}^{\sigma}\left(\varepsilon\right)\right\} \nonumber \\
 & + & \left[q_{j\sigma}\left|A_{l\sigma}\right|\right.\sin\left(\alpha_{l\sigma}+\frac{\pi}{2}\right)+q_{l\sigma}\left|A_{j\sigma}\right|\nonumber \\
 & \times & \left.\sin\left(\alpha_{j\sigma}+\frac{\pi}{2}\right)+\left|A_{j}\right|\left|A_{l}\right|\sin\left(\alpha_{j\sigma}-\alpha_{l\sigma}\right)\right]\nonumber \\
 & \times & {\tt Re}\left\{ \tilde{\mathcal{G}}_{d_{j}d_{l}}^{\sigma}\left(\varepsilon\right)\right\} .\label{eq:LDOS_p2}
\end{eqnarray}

The set of Eqs. (\ref{eq:LDOS_p1}) and (\ref{eq:LDOS_p2}) is the
main analytical finding of this work. It describes the spin-dependent
LDOS in ferromagnetic hosts with two impurities localized at distinct sites
$\vec{R}_{j}$. In the absence of the last term of Eq. (\ref{eq:LDOS_p1}),
it reduces to the case of two decoupled systems with one impurity
each. The terms in Eq. (\ref{eq:LDOS_p2}), indeed, hybridize such
single-impurity problems via the mixed Green functions $\tilde{\mathcal{G}}_{d_{j}d_{l}}^{\sigma}\left(\varepsilon\right)$.
Thus the LDOS formula encodes the single Fano factor $q_{j\sigma}$,
the Friedel-like function $A_{j\sigma}$ and the new interfering term
$\Delta\varrho_{jl\sigma}$.

We close this section by recalling that the phase $\alpha_{j\sigma}$
is nonzero for the quantum wire system, as we shall see later. The quantities
$q_{j\sigma}$ and $A_{j\sigma}$ in the quantum wire device are indeed functions
that exhibit undamped oscillations as the tip moves away from the impurities.
Conversely, damped oscillations are predicted in the metallic surface setup. In this case $\alpha_{j\sigma}=0$ and
$A_{j\sigma}$ becomes a real function. Thus the quantity $\left|A_{j\sigma}\right|$ should be read just as $A_{j\sigma}$ in Eqs. (\ref{eq:LDOS_p1}) and (\ref{eq:LDOS_p2}).} Moreover,
a ferromagnetic environment is characterized by two spin-dependent Fermi wave
numbers, namely, $k_{F\uparrow}$ and $k_{F\downarrow}$, which at
low polarization $P$ introduces a slightly difference between them.
As a result, this feature leads to a full LDOS $\rho_{LDOS}^{\uparrow}+\rho_{LDOS}^{\downarrow}$
with spin-polarized quantum beats, that can be damped or undamped, depending upon the
system dimensionality. We shall look more closely at these features later.

In STM experiments, in particular within the linear response regime
and neglecting tip-adatom coupling, the differential conductance $G=G^{\uparrow}+G^{\downarrow}$
is the observable measured by the tip, whose spin component is given
by\cite{Seridonio}

\begin{equation}
G^{\sigma}=\frac{e^{2}}{h}\pi\Gamma_{tip}\int_{-\infty}^{+\infty}\rho_{LDOS}^{\sigma}\left(\varepsilon\right)\left[-\frac{\partial f}{\partial\varepsilon}\left(\varepsilon-\phi\right)\right]d\varepsilon,\label{eq:DC}
\end{equation}
 where $e$ is the electron charge ($e>0$), $\Gamma_{tip}$ is the tip-host
coupling, $f$ is the Fermi-Dirac distribution and $\phi$ is the applied bias. For $\phi<0$ the host is the source of electrons and the tip is the drain. For $\phi>0$, we have the opposite. It is useful to define the dimensionless LDOS

\begin{equation}
LDOS=\frac{\rho_{LDOS}^{\uparrow}+\rho_{LDOS}^{\downarrow}}{\rho_{FM\uparrow}+\rho_{FM\downarrow}}\label{eq:a_LDOS}
\end{equation}
 and the transport polarization,
\begin{equation}
P_{T}=\frac{G^{\uparrow}-G{}^{\downarrow}}{G^{\uparrow}+G^{\downarrow}},\label{eq:SP_full}
\end{equation}
in order to investigate the spin-polarized quantum beats as we shall see in Sec \ref{sec4}.
Recall that, in the absence of the impurities, the transport
polarization of Eq. (\ref{eq:SP_full}) becomes $P_{T}=P$, as established
by Eq. (\ref{eq:SP}). In Sec. \ref{sec4}, we shall verify that Eq.
(\ref{eq:SP_full}) oscillates around $P$, exhibiting two distinct
behaviors as a result of the system dimensionality -- damped spin-polarized quantum beats
in the metallic surface setup and an undamped pattern in the quantum wire device.

\subsection{Fano and Friedel-like functions for the metallic surface system}

\label{sub:sec2C}

In this section, we calculate the expressions for the Fano parameter
{[}Eq. (\ref{eq:Fano_j}){]} and the Friedel-like function {[}Eq.
(\ref{eq:Friedel_j}){]} in the metallic surface case, where no manifestation
of {}``electron focusing'' occurs. This calculation was previously
performed in the single-impurity problem \cite{Seridonio} and now
we present an extension applied to the double impurity system. We
begin by solving Eq. (\ref{eq:Friedel_j}). To this end, we assume
$\phi_{\vec{k}\sigma}(\vec{R})=e^{ikRcos\theta_{k\sigma}}$
for the electronic 2D wave function and use

\begin{equation}
J_{0}\left(\xi\right)=\frac{1}{2\pi}\int_{0}^{2\pi}e^{i\xi\cos\theta_{k\sigma}}d\theta_{k\sigma},\label{eq:Jo-1}
\end{equation}
 the angular representation for the zeroth-order Bessel function.
Thus, in the wide-band limit $\left|\varepsilon\right|\ll D_{\sigma}$
with the flat-band DOS

\begin{equation}
\rho_{FM\sigma}=\frac{\mathcal{S}}{\mathcal{N}_{\sigma}2\pi}\left\{ k\left(\frac{d\varepsilon_{k\sigma}}{dk}\right)^{-1}\right\} _{k=k_{F\sigma}},\label{eq:rho_2d}
\end{equation}
 expressed in terms of the spin-dependent Fermi wave number $k_{F\sigma}$
and an element of area $\mathcal{S}$ in the host, we find

\begin{equation}
A_{j\sigma}=\frac{\rho_{FM\sigma}}{\rho_{0}}VJ_{0}\left(k_{F\sigma}\tilde{R}\right)\equiv A_{j\sigma}^{2D},\label{eq:_soma1_}
\end{equation}
 for the Friedel-like function with $\tilde{R}=\left|\vec{R}-\vec{R_{j}}\right|$
as the relative coordinate with respect to the $j$th impurity. Notice
that, according to Eq. (\ref{eq:Friedel_j}), phase $\alpha_{j\sigma}$
is zero and Eq. (\ref{eq:_soma1_}) is a real quantity. In the case
of the Fano parameter, we start defining the advanced Green function

\begin{equation}
\bar{\mathcal{G}}{}_{j\sigma}=\frac{1}{\mathcal{\mathcal{\mathcal{N}_{\sigma}}}}\sum_{\vec{k}}\frac{V_{j\vec{k}\sigma}\phi_{\vec{k}\sigma}^{*}\left(\vec{R}_{j}\right)\phi_{\vec{k}\sigma}\left(\vec{R}\right)}{\varepsilon-\varepsilon_{\vec{k}\sigma}-i\eta}\label{eq:ADV_GF}
\end{equation}
 by assuming the Lorentzian shape

\begin{equation}
V_{j\vec{k}\sigma}=V\frac{\Delta^{2}}{\Delta^{2}+\varepsilon_{k\sigma}^{2}},\label{eq:Hyb}
\end{equation}
 for an energy-dependent coupling \cite{Seridonio} in order to obtain
an analytical solution for $q_{j\sigma}.$ Notice that, in the limit
$\Delta\gg\left|\varepsilon_{k\sigma}\right|$, Eq. (\ref{eq:Hyb})
recovers the case $V_{j\vec{k}\sigma}=V$. Thus we can write the
identities
\begin{equation}
q_{j\sigma}=\frac{1}{\pi\rho_{0}}{\tt Re}\left\{ \bar{\mathcal{G}}{}_{j\sigma}\right\} \equiv q_{j\sigma}^{2D}\label{eq:real}
\end{equation}
 and

\begin{equation}
A_{j\sigma}^{2D}=\frac{1}{\pi\rho_{0}}{\tt Im}\left\{ \bar{\mathcal{G}}{}_{j\sigma}\right\} ,\label{eq:imag}
\end{equation}
 which allow us to close the calculation. Equations ~(\ref{eq:real}) and
(\ref{eq:imag}) imply the relation

\begin{equation}
q_{j\sigma}^{2D}=\frac{1}{\pi\rho_{0}}\bar{\mathcal{G}}{}_{j\sigma}-iA_{j\sigma}^{2D}.\label{eq:Fano_cb_II}
\end{equation}

As the Friedel function is already known from Eq. (\ref{eq:_soma1_}),
the quantity $\frac{1}{\pi\rho_{0}}\bar{\mathcal{G}}{}_{j\sigma}\left(\varepsilon,R\right)$
provides a relationship for the Fano parameter. To this end, we can
write
\begin{equation}
\frac{1}{\pi\rho_{0}}\bar{\mathcal{G}}{}_{j\sigma}=\frac{1}{2}\frac{\rho_{FM\sigma}}{\rho_{0}}V\sum_{l=1}^{2}\bar{\mathcal{G}}{}_{jl\sigma},\label{eq:GI}
\end{equation}
 with $\bar{\mathcal{G}}{}_{jl\sigma}\left(\varepsilon,R\right)$
obeying the following integral representation:
\begin{eqnarray}
\bar{\mathcal{G}}{}_{jl\sigma} & =\frac{1}{\pi} & \int_{-\infty}^{+\infty}d\varepsilon_{k\sigma}\frac{\Delta^{2}}{\Delta^{2}+\varepsilon_{k\sigma}^{2}}\frac{1}{\varepsilon-\varepsilon_{k\sigma}-i\eta}\nonumber \\
 & \times & H_{0}^{\left(l\right)}\left(k_{\sigma}\tilde{R}\right).\label{eq:I_1}
\end{eqnarray}

In the equation above we have used, for the sake of simplicity, the
linear dispersion relation
\begin{equation}
\varepsilon_{k\sigma}=D_{\sigma}k_{F\sigma}^{-1}\left(k-k_{F\sigma}\right),\label{eq:energy}
\end{equation}
 the Hankel functions $H_{0}^{\left(1\right)}(\xi)=J_{0}(\xi)+iY_{0}(\xi)$
and $H_{0}^{\left(2\right)}(\xi)=J_{0}(\xi)-iY_{0}(\xi)$. We remark
that Eq.~(\ref{eq:energy}) in combination with Eq.~(\ref{eq:rho_spin})
for the ferromagnetic host allows us to find

\begin{equation}
k_{F\uparrow}=\sqrt{\frac{1-P}{1+P}}k_{F\downarrow}.\label{eq:kFs}
\end{equation}
 In particular, for a small polarization $P$, Eq.~(\ref{eq:kFs})
results in slightly different Fermi wave numbers and, consequently,
in spin-polarized quantum beats in the full LDOS as we shall see. Looking at Eq. (\ref{eq:I_1}),
we calculate the integral $\bar{\mathcal{G}}{}_{j1\sigma}$ by choosing a counterclockwise contour over a semicircle in the upper-half
of the complex plane, which includes the simple pole $\varepsilon_{k\sigma}=+i\Delta$.
Applying the residue theorem, we obtain
\begin{equation}
\bar{\mathcal{G}}{}_{j1\sigma}=H_{0}^{\left(1\right)}\left(k_{\Delta}\tilde{R}\right)\frac{\Delta}{\varepsilon-i\Delta},\label{eq:I1}
\end{equation}
 with $k_{\Delta}=k_{F\sigma}(1+i\frac{\Delta}{D_{\sigma}})$.
For the evaluation of $\bar{\mathcal{G}}{}_{j2\sigma}$ we used a
clockwise contour over a semicircle in the lower-half plane, including
the poles $\varepsilon_{k\sigma}=\varepsilon-i\eta$ and $\varepsilon_{k\sigma}=-i\Delta$,
which yields
\begin{eqnarray}
\bar{\mathcal{G}}{}_{j2\sigma} & = & 2iH_{0}^{\left(2\right)}\left(k_{\varepsilon}\tilde{R}\right)\frac{\Delta^{2}}{\Delta^{2}+\varepsilon^{2}}\nonumber \\
 & + & \frac{\Delta}{\varepsilon+i\Delta}H_{0}^{\left(2\right)}\left(k_{\Delta}^{*}\tilde{R}\right),\label{eq:I2}
\end{eqnarray}
 with $k_{\varepsilon}=k_{F\sigma}\left(1+\frac{\varepsilon}{D_{\sigma}}\right).$
Taking into account the property $H_{0}^{\left(2\right)}\left(\xi\right)=\left[H_{0}^{\left(1\right)}\left(\xi^{*}\right)\right]^{*}$
for the second term in Eq. (\ref{eq:I2}), Eq. (\ref{eq:GI}) becomes

\begin{align}
\frac{1}{\pi\rho_{0}}\bar{\mathcal{G}}{}_{j\sigma} & =\frac{\rho_{FM\sigma}}{\rho_{0}}V\left[iH_{0}^{\left(2\right)}\left(k_{\varepsilon}\tilde{R}\right)\frac{\Delta^{2}}{\Delta^{2}+\varepsilon^{2}}\right.\nonumber \\
 & +\left.{\tt Re}\left\{ H_{0}^{\left(1\right)}\left(k_{\Delta}\tilde{R}\right)\frac{\Delta}{\varepsilon-i\Delta}\right\} \right].\label{eq:G_I2}
\end{align}
Explicit calculation of the terms in the brackets of Eq.~(\ref{eq:G_I2})
leads to
\begin{align}
iH_{0}^{\left(2\right)}\left(k_{\varepsilon}\tilde{R}\right)\frac{\Delta^{2}}{\Delta^{2}+\varepsilon^{2}} = iJ_{0}\left(k_{F\sigma}\tilde{R}\right)+Y_{0}\left(k_{F\sigma}\tilde{R}\right)\label{eq:ft}
\end{align}
 and
\begin{align}
{\tt Re}\left\{ H_{0}^{\left(1\right)}\left(k_{\Delta}\tilde{R}\right)\frac{\Delta}{\varepsilon-i\Delta}\right\} = - Y_{0}\left(k_{F\sigma}\tilde{R}\right),\label{eq:st}
\end{align}
 where we have assumed $\left|\varepsilon\right|\ll D_{\sigma}$,
$\Delta\ll D_{\sigma}$ and $\Delta\gg\left|\varepsilon\right|$. In order to ensure
the limit $V_{j\vec{k}\sigma}=V$ in Eq. (\ref{eq:Hyb}), we make the substitution
of Eqs. (\ref{eq:_soma1_}), (\ref{eq:G_I2}), (\ref{eq:ft}) and
(\ref{eq:st}) in Eq. (\ref{eq:Fano_cb_II}), showing that

\begin{align}
q_{j\sigma}^{2D} = 0\label{eq:Fano_j_2}
\end{align}
for any value of $k_{F\sigma}\tilde{R}$.

In summary, the zero value of the Fano parameter given by Eq. (\ref{eq:Fano_j_2})
and the zeroth-order Bessel function $J_{0}\left(k_{F\sigma}\tilde{R}\right)$ found in Eq. (\ref{eq:_soma1_}) lead to long-range
damped spin-polarized quantum beats in the full LDOS. This feature will be discussed in Sec. \ref{sec4}.

\subsection{Fano and Friedel-like functions for the quantum wire system}

\label{sub:sec2D}

Here we determine the Fano parameter in Eq. (\ref{eq:Fano_j}) and
the Friedel-like function in Eq. (\ref{eq:Friedel_j}) for the quantum wire
case. Following A. Weismann,\cite{STM20_FS} we use $\phi_{\vec{k}\sigma}(\vec{R})=e^{ikR}$
as the electron wave function, in which the direction introduced by
$\vec{R}$ defines the STM tip-impurity distance where {}``electron
focusing'' manifests. We also use the dispersion relation

\begin{equation}
\varepsilon_{k\sigma}=\frac{\hbar^{2}k^{2}}{2m}-D_{\sigma}\label{eq:dispersion_relation}
\end{equation}
 and the flat DOS

\begin{equation}
\rho_{FM\sigma}=\frac{\mathcal{L}}{\mathcal{N}_{\sigma}2\pi}\left(\frac{d\varepsilon_{k\sigma}}{dk}\right)_{k=k_{F\sigma}}^{-1}=\frac{\mathcal{L}}{\mathcal{N}_{\sigma}2\pi}\frac{m}{\hbar^{2}k_{F\sigma}},\label{eq:QW_f_LDOS}
\end{equation}
 with $m$ as the effective electron mass and $\mathcal{L}$ as a
given length in the wire. Additionally, in the wide-band limit $\left|\varepsilon\right|\ll D_{\sigma}$,
we find the following complex Friedel-like function:

\begin{align}
A_{j\sigma} & =V\frac{\rho_{FM\sigma}}{\rho_{0}}e^{ik_{F\sigma}\tilde{R}}\equiv A_{j\sigma}^{1D},\label{eq:soma_2}
\end{align}
 characterized by a spin-dependent phase $\alpha_{j\sigma}=k_{F\sigma}\tilde{R}$.
This phase results in nondamped, oscillatory behavior as a function
of $\tilde{R}$, which also appears in $q_{j\sigma}^{1D}$. Thus we take into account Eqs. (\ref{eq:dispersion_relation}) and (\ref{eq:QW_f_LDOS}),
rewriting Eq. (\ref{eq:Fano_j}) as

\begin{equation}
q_{j\sigma}=\frac{V}{\pi\rho_{0}\mathcal{\mathcal{\mathcal{N}_{\sigma}}}}\sum_{\vec{k}}\frac{e^{ik\tilde{R}}}{\varepsilon-\varepsilon_{k}}=2V\frac{\mathcal{L}}{\rho_{0}\mathcal{N}_{\sigma}\pi}\frac{m}{\hbar^{2}}\mathcal{I}\equiv q_{j\sigma}^{1D},\label{eq:pre_c1}
\end{equation}
 with
\begin{equation}
\mathcal{I}=-\frac{1}{2\pi}\wp\int_{-\infty}^{+\infty}\frac{e^{ik\tilde{R}}}{k^{2}-k_{\varepsilon\sigma}^{2}}dk=\frac{\sin\left(k_{F\sigma}\tilde{R}\right)}{2k_{F\sigma}}\label{eq:pre_c2}
\end{equation}
 in the limit $\left|\varepsilon\right|\ll D_{\sigma}$ and with $\wp$
as the principal value. Finally, we obtain the Fano factor

\begin{align}
q_{j\sigma}^{1D} & =2V\frac{\rho_{FM\sigma}}{\rho_{0}}\sin\left(k_{F\sigma}\tilde{R}\right),\label{eq:fano_1d}
\end{align}
 which also presents spin-dependent Fermi wave numbers $k_{F\uparrow}$
and $k_{F\downarrow}$ as in Eq.~(\ref{eq:soma_2}). Here they are
still connected via Eq.~(\ref{eq:kFs}), thus leading to undamped
spin-polarized quantum beats in the full LDOS.

\section{CALCULATION OF THE IMPURITY GREEN FUNCTION}

\label{sec3}

In the present section we calculate $\tilde{\mathcal{G}}_{d_{j}d_{l}}^{\sigma}\left(\varepsilon\right)$
($j,l=1,2$) that appear in Eqs. (\ref{eq:LDOS_p1}) and (\ref{eq:LDOS_p2})
for the LDOS. To handle the interacting term of the Hamiltonian, we
adopt the Hubbard I approximation,\cite{book2} which provides reliable
results at temperatures above the Kondo temperature. \cite{book2}
We begin by repeating the equation-of-motion approach for these Green functions, which results
in

\begin{eqnarray}
\left(\varepsilon-\tilde{\varepsilon}_{jd\sigma}+i\Gamma_{jj\sigma}\right)\tilde{\mathcal{G}}_{d_{j}d_{j}}^{\sigma} & = & 1+U_{j}\tilde{\mathcal{G}}_{d_{j\sigma}n_{d_{j}\bar{\sigma}},d_{j\sigma}}\nonumber \\
+\sum_{l\neq j}\left(\Sigma_{lj\sigma}^{R}-i\Gamma_{lj\sigma}\right) & \times & \tilde{\mathcal{G}}_{d_{l}d_{j}}^{\sigma}\label{eq:s1}
\end{eqnarray}
 and
\begin{align}
\left(\varepsilon-\tilde{\varepsilon}_{ld\sigma}+i\Gamma_{ll\sigma}\right)\tilde{\mathcal{G}}_{d_{l}d_{j}}^{\sigma} & =U_{l}\tilde{\mathcal{G}}_{d_{l\sigma}n_{d_{l}\bar{\sigma}},d_{j\sigma}}\nonumber \\
+\left(\Sigma_{jj\sigma}^{R}-i\Gamma_{jj\sigma}\right) & \times\tilde{\mathcal{G}}_{d_{j}d_{j}}^{\sigma},\label{eq:s2}
\end{align}
 where $\tilde{\varepsilon}_{jd\sigma}=\varepsilon_{jd\sigma}+\Sigma_{jj\sigma}^{R}$
for $l\neq j$. In the equation above, $\tilde{\mathcal{G}}_{d_{l\sigma}n_{d_{l}\bar{\sigma}},d_{j\sigma}}$
is a higher-order Green function obtained from the time Fourier transform of
\begin{align}
\mathcal{G}_{d_{l\sigma}n_{d_{l}\bar{\sigma}},d_{j\sigma}}\left(t\right) & =-\frac{i}{\hbar}\theta\left(t\right)\mathcal{Z}_{FM}^{-1}\sum_{n}e^{-\beta E_{n}}\nonumber \\
 & \times\left\langle n\right|\left[d_{l\sigma}\left(t\right)n_{d_{l}\bar{\sigma}}\left(t\right),d_{j\sigma}^{\dagger}\left(0\right)\right]_{+}\left|n\right\rangle ,\label{eq:H_GF}
\end{align}
 with $n_{d_{l}\bar{\sigma}}=d_{l\bar{\sigma}}^{\dagger}d_{l\bar{\sigma}}$
being the number operator of the $l$th impurity with spin $\bar{\sigma}$
(opposite to $\sigma$). Here
\begin{equation}
\Sigma_{lj\sigma}^{R}=\frac{1}{\mathcal{N}_{\sigma}}\sum_{\vec{k}}\frac{V_{j\vec{k}\sigma}^{*}V_{l\vec{k}\sigma}\phi_{\vec{k}\sigma}\left(\vec{R}_{j}\right)\phi_{\vec{k}\sigma}^{*}\left(\vec{R}_{l}\right)}{\varepsilon-\varepsilon_{\vec{k}\sigma}}\label{eq:r_s_e}
\end{equation}
 represents the real part of the noninteracting self-energy $\Sigma_{lj\sigma}$
and
\begin{align}
\Sigma_{lj\sigma}^{I} & =-\Gamma_{lj\sigma}=-\frac{1}{\mathcal{N}_{\sigma}}\pi\sum_{\vec{k}}V_{j\vec{k}\sigma}^{*}V_{l\vec{k}\sigma}\phi_{\vec{k}\sigma}\left(\vec{R}_{j}\right)\phi_{\vec{k}\sigma}^{*}\left(\vec{R}_{l}\right)\nonumber \\
 & \times\delta\left(\varepsilon-\varepsilon_{\vec{k}\sigma}\right)\label{eq:i_s_e}
\end{align}
 describes the corresponding imaginary part, which plays the role
of a generalized Anderson parameter $\Gamma_{lj\sigma}$. In order
to close the system of Green functions in Eqs. (\ref{eq:s1}) and (\ref{eq:s2}),
we first take the time derivative of Eq. (\ref{eq:H_GF}) and then
perform the time Fourier transform. With that we obtain

\begin{eqnarray}
\left(\varepsilon^{+}-\varepsilon_{ld\sigma}-U_{l}\right)\tilde{\mathcal{G}}_{d_{l\sigma}n_{d_{l}\bar{\sigma}},d_{j\sigma}} & = & \delta_{lj}\left\langle n_{d_{l}\bar{\sigma}}\right\rangle \nonumber \\
+\left(-\sum_{\vec{k}}\frac{1}{\sqrt{\mathcal{N}_{\bar{\sigma}}}}V_{l\vec{k}\sigma}\phi_{\vec{k}\bar{\sigma}}^{*}\left(\vec{R}_{l}\right)\right) & \times & \tilde{\mathcal{G}}_{c_{\vec{k}\bar{\sigma}}^{\dagger}d_{l\bar{\sigma}}d_{l\sigma},d_{j\sigma}}\nonumber \\
+\left(\sum_{\vec{k}}\frac{1}{\sqrt{\mathcal{N}_{\sigma}}}V_{l\vec{k}\sigma}^{*}\phi_{\vec{k}\sigma}\left(\vec{R}_{l}\right)\right) & \times & \tilde{\mathcal{G}}_{c_{\vec{k}\sigma}d_{l\bar{\sigma}}^{\dagger}d_{l\bar{\sigma}},d_{j\sigma}}\nonumber \\
+\left(\sum_{\vec{k}}\frac{1}{\sqrt{\mathcal{N}_{\bar{\sigma}}}}V_{l\vec{k}\sigma}^{*}\phi_{\vec{k}\bar{\sigma}}\left(\vec{R}_{l}\right)\right) & \times & \tilde{\mathcal{G}}_{d_{l\bar{\sigma}}^{\dagger}c_{\vec{k}\bar{\sigma}}d_{l\sigma},d_{j\sigma}},\nonumber \\
\label{eq:H_GF_2}
\end{eqnarray}
 which also depends on new Green functions of the same order of $\tilde{\mathcal{G}}_{d_{l\sigma}n_{d_{l}\bar{\sigma}},d_{j\sigma}}$
and on the average occupation number $\left\langle n_{d_{l}\bar{\sigma}}\right\rangle $,
that is, calculated as
\begin{equation}
\left\langle n_{d_{l}\bar{\sigma}}\right\rangle =\int_{-\infty}^{+\infty}d\varepsilon\left\{ -\frac{1}{\pi}{\tt Im}\left(\tilde{\mathcal{G}}_{d_{l}d_{l}}^{\bar{\sigma}}\right)\right\} f\left(\varepsilon\right).\label{eq:nb}
\end{equation}

Within the Hubbard I approximation we truncate the Green functions $\tilde{\mathcal{G}}_{c_{\vec{k}\bar{\sigma}}^{\dagger}d_{l\bar{\sigma}}d_{l\sigma},d_{j\sigma}}$
and $\tilde{\mathcal{G}}_{d_{l\bar{\sigma}}^{\dagger}c_{\vec{k}\bar{\sigma}}d_{l\sigma},d_{j\sigma}}$
according to the decoupling scheme

\begin{align}
\tilde{\mathcal{G}}_{c_{\vec{k}\bar{\sigma}}^{\dagger}d_{l\bar{\sigma}}d_{l\sigma},d_{j\sigma}} & \simeq\left\langle c_{\vec{k}\bar{\sigma}}^{\dagger}d_{l\bar{\sigma}}\right\rangle \tilde{\mathcal{G}}_{d_{l}d_{j}}^{\sigma}\label{eq:dc_1}\\
\tilde{\mathcal{G}}_{d_{l\bar{\sigma}}^{\dagger}c_{\vec{k}\bar{\sigma}}d_{l\sigma},d_{j\sigma}} & \simeq\left\langle c_{\vec{k}\bar{\sigma}}^{\dagger}d_{l\bar{\sigma}}\right\rangle \tilde{\mathcal{G}}_{d_{l}d_{j}}^{\sigma},\label{eq:dc_2}
\end{align}
 and apply the equation-of-motion approach to $\tilde{\mathcal{G}}_{c_{\vec{k}\sigma}d_{l\bar{\sigma}}^{\dagger}d_{l\bar{\sigma}},d_{j\sigma}}.$
In order to cancel the second term with the last one in Eq. (\ref{eq:H_GF_2}),
we combine the approximations in Eqs. (\ref{eq:dc_1}) and (\ref{eq:dc_2})
simultaneously with the property

\begin{equation}
\sum_{\vec{k}}\frac{1}{\sqrt{\mathcal{N}_{\bar{\sigma}}}}V_{l\vec{k}\sigma}\phi_{\vec{k}\bar{\sigma}}^{*}\left(\vec{R}_{l}\right)=\sum_{\vec{k}}\frac{1}{\sqrt{\mathcal{N}_{\bar{\sigma}}}}V_{l\vec{k}\sigma}^{*}\phi_{\vec{k}\bar{\sigma}}\left(\vec{R}_{l}\right),\label{eq:proper}
\end{equation}
 which is only fulfilled on a metallic surface system, while for the quantum wire, it holds
in the side-coupled geometry $\vec{R}_{l}=\vec{R}_{j}=\vec{0}$. Bearing
this in mind, we can rewrite Eq. (\ref{eq:H_GF_2}) as follows

\begin{eqnarray}
\left(\varepsilon-\varepsilon_{ld\sigma}-U_{l}+i\eta\right)\tilde{\mathcal{G}}_{d_{l\sigma}n_{d_{l}\bar{\sigma}},d_{j\sigma}} & = & \delta_{lj}\left\langle n_{d_{l}\bar{\sigma}}\right\rangle \nonumber \\
+\left(\sum_{\vec{k}}\frac{1}{\sqrt{\mathcal{N}_{\sigma}}}V_{l\vec{k}\sigma}^{*}\phi_{\vec{k}\sigma}\left(\vec{R}_{l}\right)\right) & \times & \tilde{\mathcal{G}}_{c_{\vec{k}\sigma}d_{l\bar{\sigma}}^{\dagger}d_{l\bar{\sigma}},d_{j\sigma}}.\label{eq:H_GF_3}
\end{eqnarray}
 Once again, employing the equation-of-motion approach for $\tilde{\mathcal{G}}_{c_{\vec{k}\sigma}d_{l\bar{\sigma}}^{\dagger}d_{l\bar{\sigma}},d_{j\sigma}}$,
we find

\begin{align}
\left(\varepsilon^{+}-\varepsilon_{\vec{k}\sigma}\right)\tilde{\mathcal{G}}_{c_{\vec{k}\sigma}d_{l\bar{\sigma}}^{\dagger}d_{l\bar{\sigma}},d_{j\sigma}} & =\nonumber \\
V_{l\vec{k}\sigma}\frac{1}{\sqrt{\mathcal{N}_{\sigma}}} & \times\phi_{\vec{k}\sigma}^{*}\left(\vec{R}_{l}\right)\tilde{\mathcal{G}}_{d_{l}n_{d_{l}\bar{\sigma}},d_{j\sigma}}\nonumber \\
+\sum_{\vec{q}}V_{l\vec{q}\sigma}^{*}\frac{1}{\sqrt{\mathcal{N}_{\bar{\sigma}}}} & \times\phi_{\vec{q}\bar{\sigma}}\left(\vec{R}_{l}\right)\tilde{\mathcal{G}}_{c_{\vec{k}\sigma}d_{l\bar{\sigma}}^{\dagger}c_{\vec{q}\bar{\sigma}},d_{j\sigma}}\nonumber \\
-\sum_{\vec{q}}V_{l\vec{q}\sigma}\frac{1}{\sqrt{\mathcal{N}_{\bar{\sigma}}}} & \times\phi_{\vec{q}\bar{\sigma}}^{*}\left(\vec{R}_{l}\right)\tilde{\mathcal{G}}_{c_{\vec{q}\bar{\sigma}}^{\dagger}d_{l\bar{\sigma}}c_{\vec{k}\sigma},d_{j\sigma}}\nonumber \\
+\sum_{\tilde{j}\neq l}V_{\tilde{j}\vec{k}\sigma}\frac{1}{\sqrt{\mathcal{N}_{\sigma}}} & \times\phi_{\vec{k}\sigma}^{*}\left(\vec{R}_{\tilde{j}}\right)\tilde{\mathcal{G}}_{d_{\tilde{j}\sigma}n_{d_{l}\bar{\sigma}},d_{j\sigma}}.\label{eq:H_GF_4}
\end{align}

Here we continue with the Hubbard I scheme, proceeding as in Eqs.
(\ref{eq:dc_1}) and (\ref{eq:dc_2}) by making the following approximations:

\begin{align}
\tilde{\mathcal{G}}_{c_{\vec{k}\sigma}d_{l\bar{\sigma}}^{\dagger}c_{\vec{q}\bar{\sigma}},d_{j\sigma}} & \simeq\left\langle d_{l\bar{\sigma}}^{\dagger}c_{\vec{q}\bar{\sigma}}\right\rangle \tilde{\mathcal{G}}_{c_{\vec{k}\sigma}d_{j\sigma}},\label{eq:dc_3}\\
\tilde{\mathcal{G}}_{c_{\vec{q}\bar{\sigma}}^{\dagger}d_{l\bar{\sigma}}c_{\vec{k}\sigma},d_{j\sigma}} & \simeq\left\langle d_{l\bar{\sigma}}^{\dagger}c_{\vec{q}\bar{\sigma}}\right\rangle \tilde{\mathcal{G}}_{c_{\vec{k}\sigma}d_{j\sigma},}\label{eq:dc_4}\\
\tilde{\mathcal{G}}_{d_{\tilde{j}\sigma}n_{d_{l}\bar{\sigma}},d_{j\sigma}} & \simeq\left\langle n_{d_{l}\bar{\sigma}}\right\rangle \tilde{\mathcal{G}}_{d_{\tilde{j}}d_{j}}^{\sigma}
\end{align}
 and replacing Eq. (\ref{eq:proper}) in Eq. (\ref{eq:H_GF_4}) to
show that

\begin{align}
\tilde{\mathcal{G}}_{c_{\vec{k}\sigma}d_{l\bar{\sigma}}^{\dagger}d_{l\bar{\sigma}},d_{j\sigma}} & =\frac{V_{l\vec{k}\sigma}\frac{1}{\sqrt{\mathcal{N}_{\sigma}}}\phi_{\vec{k}\sigma}^{*}\left(\vec{R}_{l}\right)}{\left(\varepsilon^{+}-\varepsilon_{\vec{k}\sigma}\right)}\tilde{\mathcal{G}}_{d_{l}n_{d_{l}\bar{\sigma}},d_{j\sigma}}\nonumber \\
 & +\frac{\sum_{\tilde{j}\neq l}V_{\tilde{j}\vec{k}\sigma}\frac{1}{\sqrt{\mathcal{N}_{\sigma}}}\phi_{\vec{k}\sigma}^{*}\left(\vec{R}_{\tilde{j}}\right)}{\left(\varepsilon^{+}-\varepsilon_{\vec{k}\sigma}\right)}\nonumber \\
 & \times\left\langle n_{d_{l}\bar{\sigma}}\right\rangle \tilde{\mathcal{G}}_{d_{\tilde{j}}d_{j}}^{\sigma}.\label{eq:HG_F5}
\end{align}

To close the original setup of Green functions in Eqs. (\ref{eq:s1}) and (\ref{eq:s2}),
we substitute Eq. (\ref{eq:HG_F5}) in Eq. (\ref{eq:H_GF_3}) and
obtain
\begin{eqnarray}
\left(\varepsilon-\varepsilon_{ld\sigma}-U_{l}+i\Gamma_{ll\sigma}^{0}\right)\tilde{\mathcal{G}}_{d_{l\sigma}n_{d_{l}\bar{\sigma}},d_{j\sigma}} & = & \delta_{lj}\left\langle n_{d_{l}\bar{\sigma}}\right\rangle \nonumber \\
+\left\langle n_{d_{l}\bar{\sigma}}\right\rangle \sum_{\tilde{j}\neq l}\left(\Sigma_{\tilde{j}l\sigma}^{R}\left(\varepsilon\right)-i\Gamma_{\tilde{j}l\sigma}\right) & \times & \tilde{\mathcal{G}}_{d_{\tilde{j}}d_{j}}^{\sigma},\label{eq:s3}
\end{eqnarray}
 where
\begin{equation}
\Gamma_{ll\sigma}^{0}=\pi\frac{1}{\mathcal{N}_{\sigma}}\sum_{\vec{k}}\left|V_{l\vec{k}\sigma}\right|^{2}\delta\left(\varepsilon-\varepsilon_{\vec{k}\sigma}\right),\label{eq:gama_zero}
\end{equation}
 which allows us to determine all the necessary Green functions for the LDOS.
Thus, to solve the system composed by Eqs. (\ref{eq:s1}), (\ref{eq:s2})
and (\ref{eq:s3}), we now assume the side-coupled geometry $\vec{R}_{l}=\vec{R}_{j}=\vec{0}$
(Fig. \ref{fig:Pic1}). A geometry with $\vec{R}_{l}\neq\vec{R}_{j}\neq\vec{0}$
will be published elsewhere. By choosing the side-coupled configuration
and assuming constant symmetric couplings $V_{j\vec{k}\sigma}=V_{l\vec{k}\sigma}=V$,
we verify from Eqs. (\ref{eq:i_s_e}) and (\ref{eq:gama_zero}) that
$\Gamma_{jj\sigma}=\Gamma_{lj\sigma}=\Gamma_{ll\sigma}^{0}=\Gamma_{\sigma}=\Gamma\frac{\rho_{FM\sigma}}{\rho_{0}}$
depends on the standard Anderson parameter $\Gamma=\pi V^{2}\rho_{0}.$
Additionally, in the wide-band limit, Eq. (\ref{eq:r_s_e})
ensures $\Sigma_{jj\sigma}^{R}=\Sigma_{lj\sigma}^{R}=0$. For the sake
of simplicity we consider the infinite Coulomb correlation limit ($U_{1}=U_{2}\rightarrow\infty$).
Thus the direct Green function for the impurity $j=1$ reduces to the form

\begin{equation}
\tilde{\mathcal{G}}_{d_{1}d_{1}}^{\sigma}\left(\varepsilon\right)=\frac{1-\left\langle n_{d_{1}\bar{\sigma}}\right\rangle }{\varepsilon-\epsilon_{12d\sigma}+i\bar{\triangle}_{12\sigma}},\label{eq:sol1}
\end{equation}
 where
\begin{equation}
\epsilon_{12d\sigma}=\varepsilon_{1d\sigma}+\bar{\Sigma}_{12\sigma}\left(\varepsilon\right)\label{
eq:sol1B}
\end{equation}
 represents a renormalized energy level dressed by the real part of
the nondiagonal self-energy
\begin{align}
\bar{\Sigma}_{12\sigma}\left(\varepsilon\right) & =-\left(1-\left\langle n_{d_{1}\bar{\sigma}}\right\rangle \right)\left(1-\left\langle n_{d_{2}\bar{\sigma}}\right\rangle \right)\nonumber \\
 & \times\frac{\left(\varepsilon-\varepsilon_{2d\sigma}\right)}{\left(\varepsilon-\varepsilon_{2d\sigma}\right)^{2}+\Gamma_{\sigma}^{2}}\Gamma_{\sigma}^{2}\label{eq:sol1C}
\end{align}
 and
\begin{align}
\bar{\triangle}_{12\sigma} & =\Gamma_{\sigma}-\left(1-\left\langle n_{d_{1}\bar{\sigma}}\right\rangle \right)\left(1-\left\langle n_{d_{2}\bar{\sigma}}\right\rangle \right)\Gamma_{\sigma}\nonumber \\
 & \times\frac{\Gamma_{\sigma}^{2}}{\left(\varepsilon-\varepsilon_{2d\sigma}\right)^{2}+\Gamma_{\sigma}^{2}}\label{eq:sol1D}
\end{align}
 is an effective hybridization function. The mixed Green function $\tilde{\mathcal{G}}_{d_{2}d_{1}}^{\sigma}\left(\varepsilon\right)$
becomes
\begin{equation}
\tilde{\mathcal{G}}_{d_{2}d_{1}}^{\sigma}\left(\varepsilon\right)=-i\Gamma_{\sigma}\left\{ \frac{1-\left\langle n_{d_{2}\bar{\sigma}}\right\rangle }{\left(\varepsilon-\varepsilon_{2d\sigma}+i\Gamma_{\sigma}\right)}\right\} \tilde{\mathcal{G}}_{d_{1}d_{1}}^{\sigma}\left(\varepsilon\right).\label{eq:sol2}
\end{equation}
 Notice that the other Green functions $\tilde{\mathcal{G}}_{d_{2}d_{2}}^{\sigma}$
and $\tilde{\mathcal{G}}_{d_{2}d_{1}}^{\sigma}$ can be derived by swapping
$1\leftrightarrow2$ in Eqs. (\ref{eq:sol1}) and (\ref{eq:sol2}).

\section{NUMERICAL RESULTS}

\label{sec4}

\subsection{Numerical Parameters}

Here we present the results obtained via the formulation developed
in the previous section. The energy scale adopted is the Anderson
parameter $\Gamma$. We employ the following set of model parameters: $\Gamma=0.2$ $eV$, $\varepsilon_{1d\sigma}=\varepsilon_{1d}=-10\Gamma$
and $\varepsilon_{2d\sigma}=\varepsilon_{2d}=-4.5\Gamma$. \cite{AHCNeto,STM13} Such values correspond to a
Kondo temperature $T_{K}\approx50K$ found in the system Co/Cu(111) with Coulomb interaction $U=2.9$ $eV$. \cite{STM6,STM11,AHCNeto} Thus the Hubbard I approximation is employed with $T=\Gamma/10k_{B}=231.1K$ just to avoid Kondo physics.
Finally, in order to generate spin-polarized quantum beats in the LDOS, we substitute Eq. (\ref{eq:SP}) with $P=0.1$ in Eq. (\ref{eq:kFs}).

\begin{figure}[h]
 \centerline{\resizebox{3.5in}{!}{ \includegraphics[clip,width=0.6\textwidth]{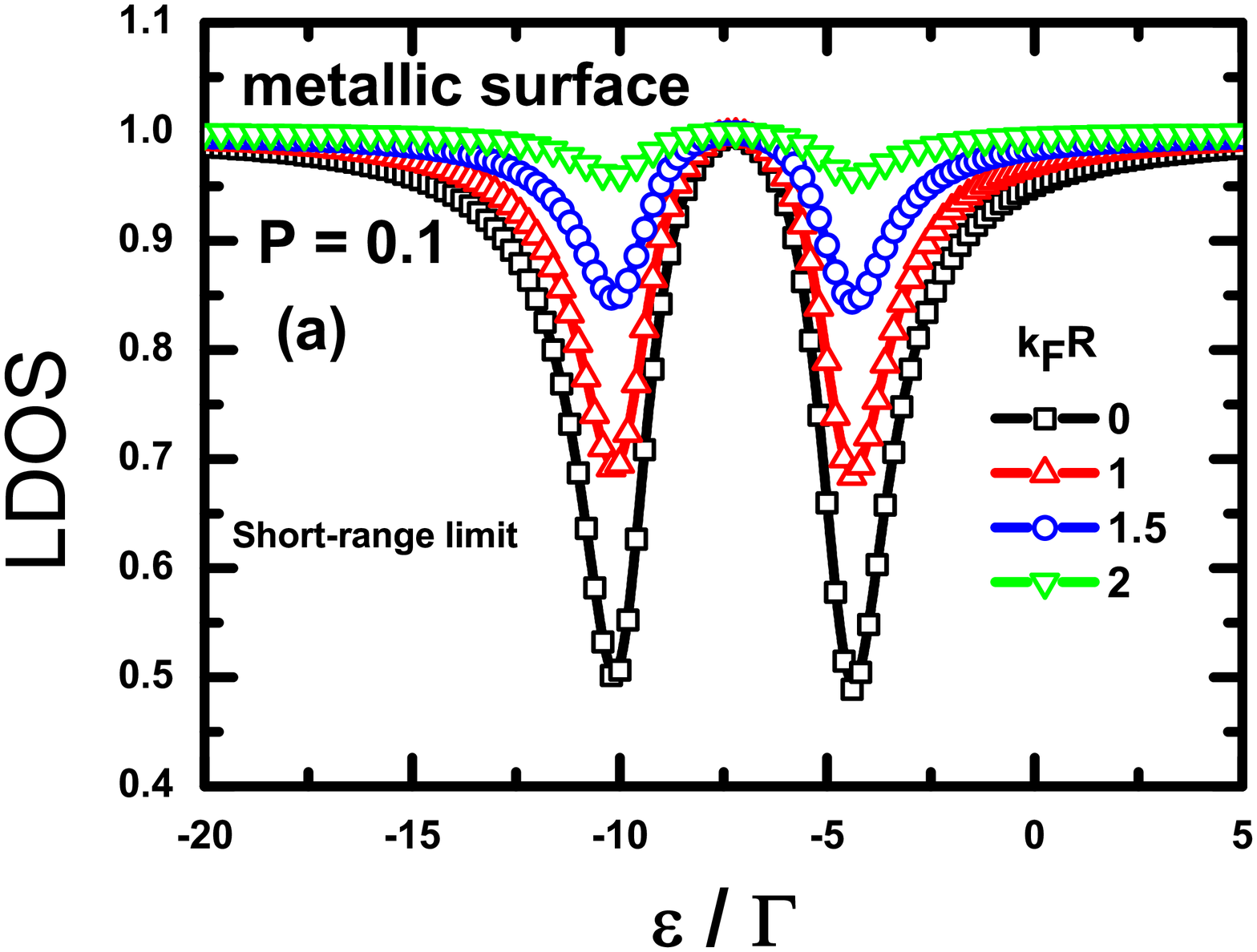}}}
\centerline{\resizebox{3.5in}{!}{ \includegraphics[clip,width=0.6\textwidth]{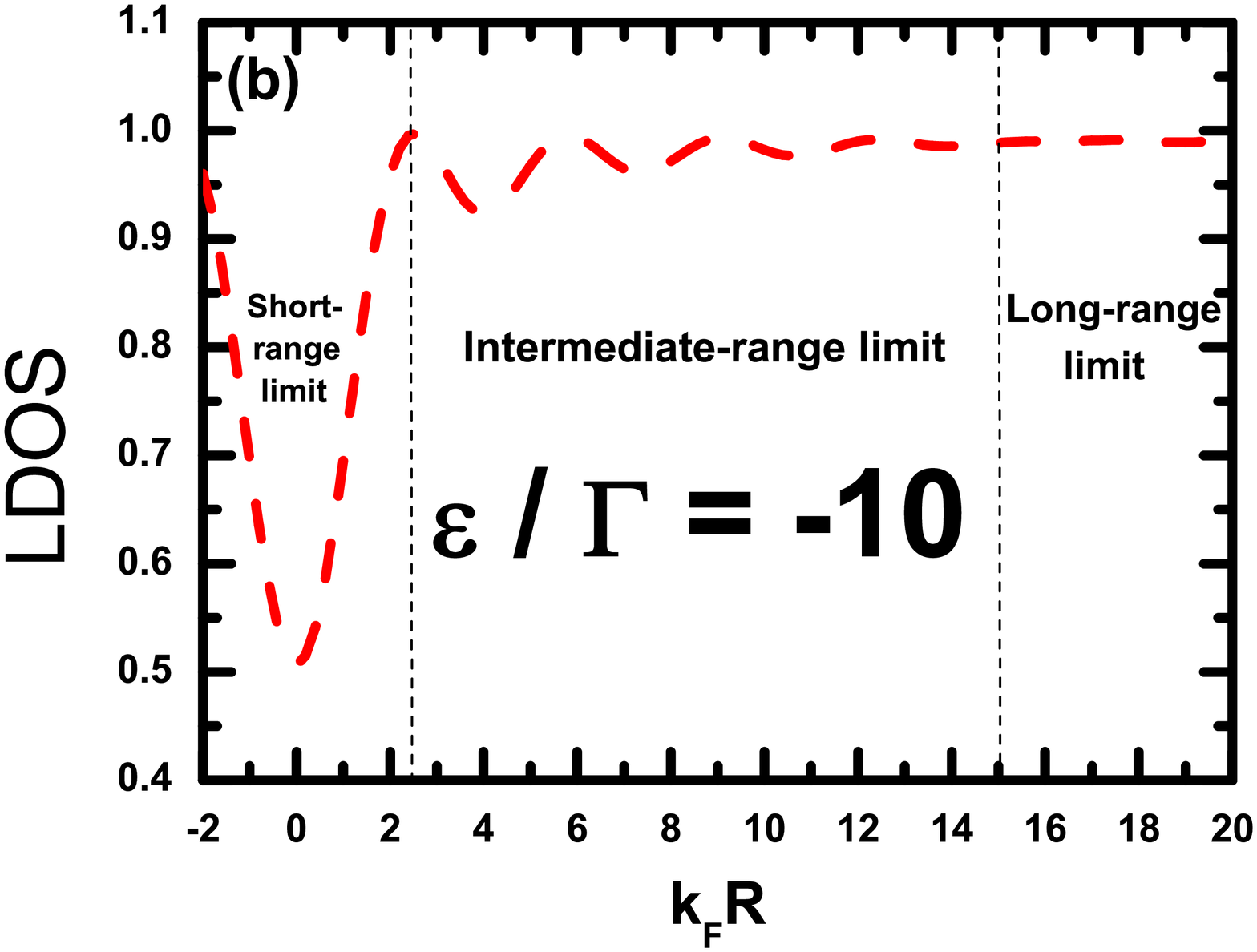}}}
\caption{(Color online) In both panels we use $k_{B}T=0.1\Gamma$. (a) LDOS {[}Eq. (\ref{eq:a_LDOS}){]}
of a metallic surface with $P=0.1$ as a function of $\varepsilon/\Gamma$ for
different values of $k_{F}R$ in the short-range limit (see panel (b)). The Fano profile
presents two antiresonances placed at $\varepsilon=\varepsilon_{1d}=-10\Gamma$
and $\varepsilon=\varepsilon_{2d}=-4.5\Gamma$, which display an evanescent
behavior for increasing distances. (b) Keeping the energy at $\varepsilon=-10\Gamma,$
Friedel oscillations appear in the LDOS. }

\label{Fig2}
\end{figure}

\begin{figure}[h]
 \centerline{\resizebox{3.5in}{!}{ \includegraphics[clip,width=0.6\textwidth]{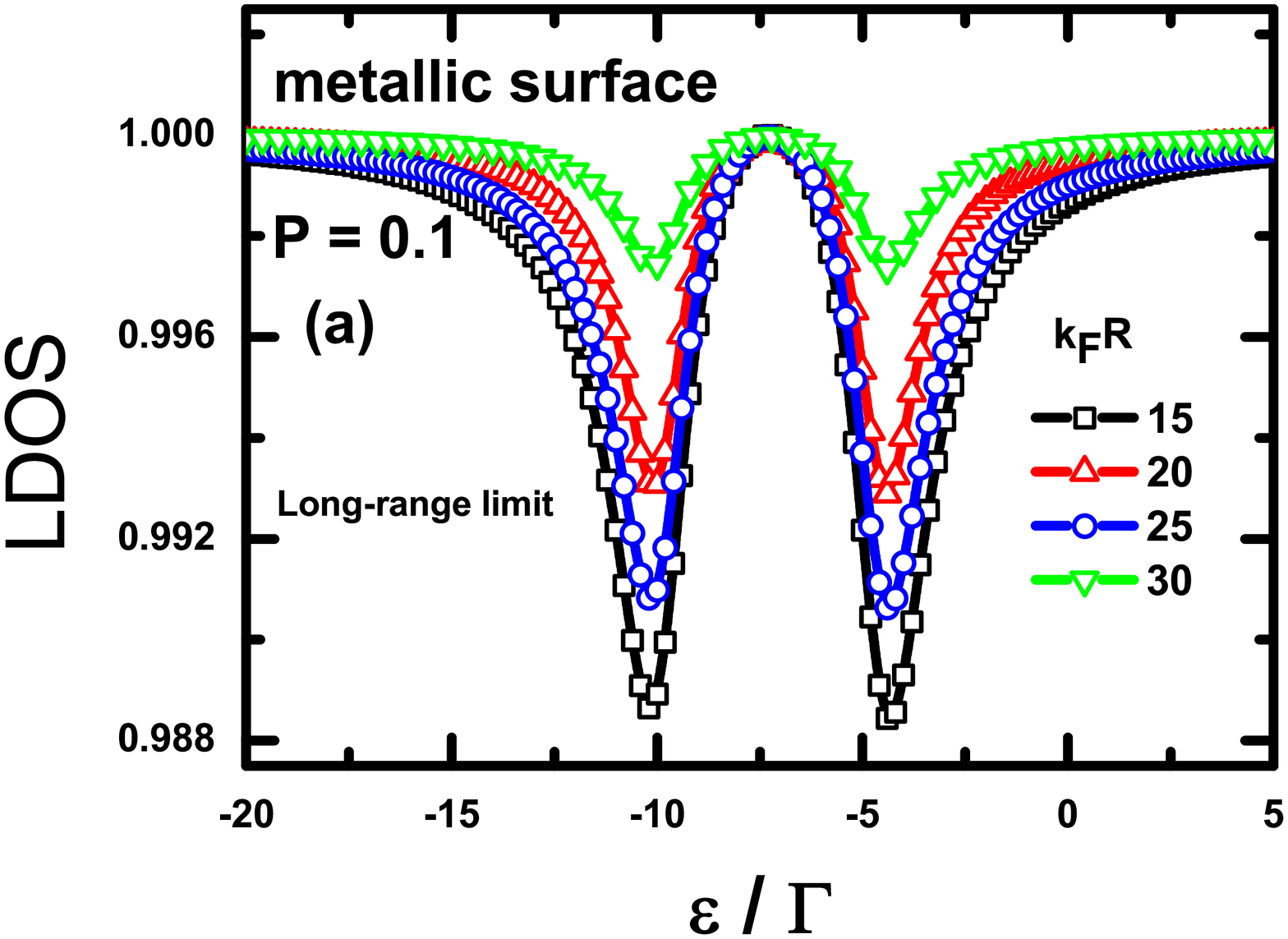}}}
\centerline{\resizebox{3.5in}{!}{ \includegraphics[clip,width=0.6\textwidth]{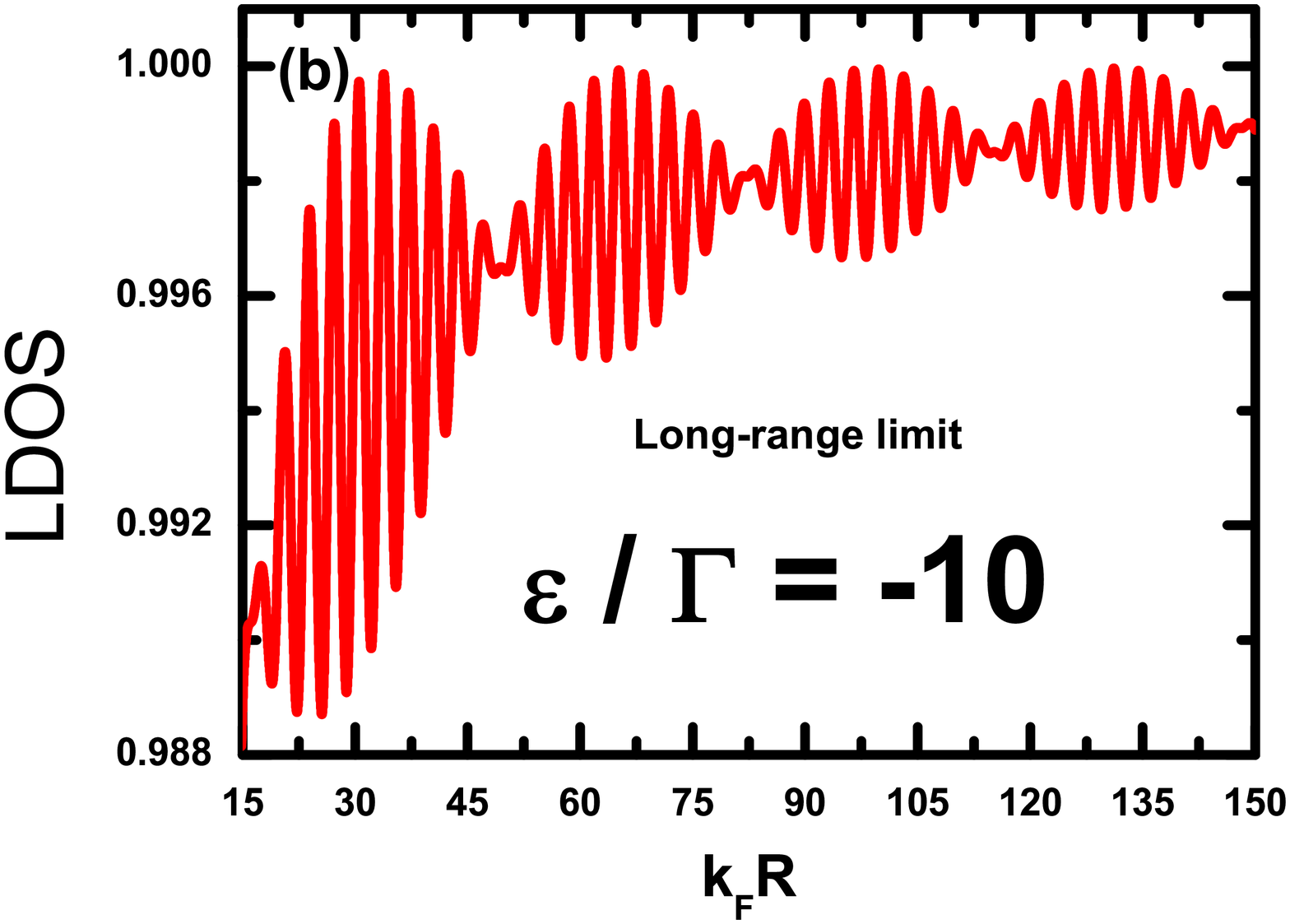}}}
\caption{(Color online) In both panels we use $k_{B}T=0.1\Gamma$. (a) LDOS {[}Eq. (\ref{eq:a_LDOS}){]}
of a metallic surface with $P=0.1$ as a function of $\varepsilon/\Gamma$ for
different values of $k_{F}R$ in the long-range limit (see panel (b)). The Fano profile
presents two antiresonances placed at $\varepsilon=\varepsilon_{1d}=-10\Gamma$
and $\varepsilon=\varepsilon_{2d}=-4.5\Gamma$, which display an oscillatory
behavior for increasing distances. (b) Damped spin-polarized quantum beats emerge in the
LDOS as function of $k_{F}R$ with $\varepsilon=\varepsilon_{1d}=-10\Gamma$. }

\label{Fig3}
\end{figure}

\begin{figure}[h]
 \centerline{\resizebox{3.5in}{!}{ \includegraphics[clip,width=0.6\textwidth]{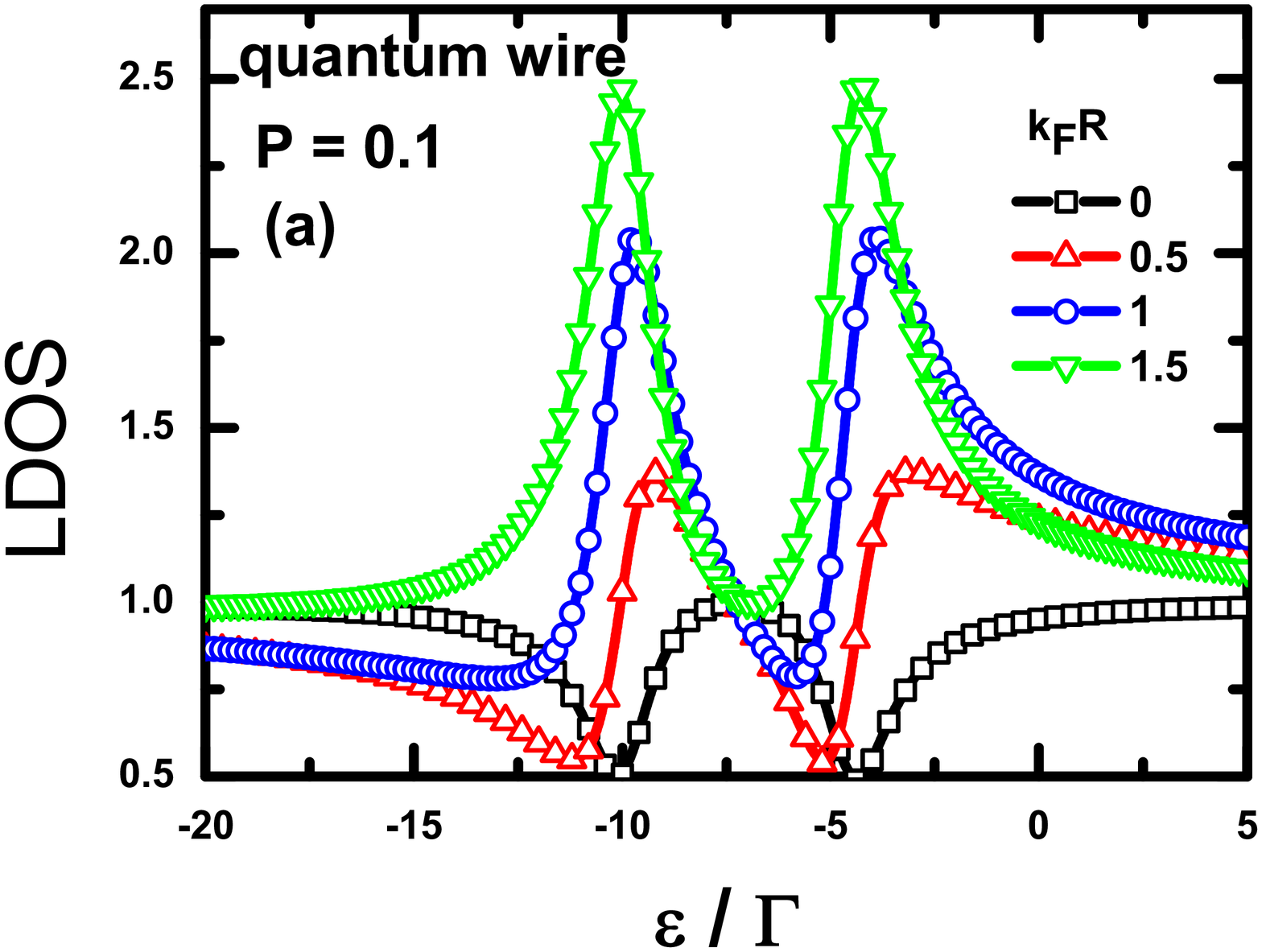}}}
\centerline{\resizebox{3.5in}{!}{ \includegraphics[clip,width=0.6\textwidth]{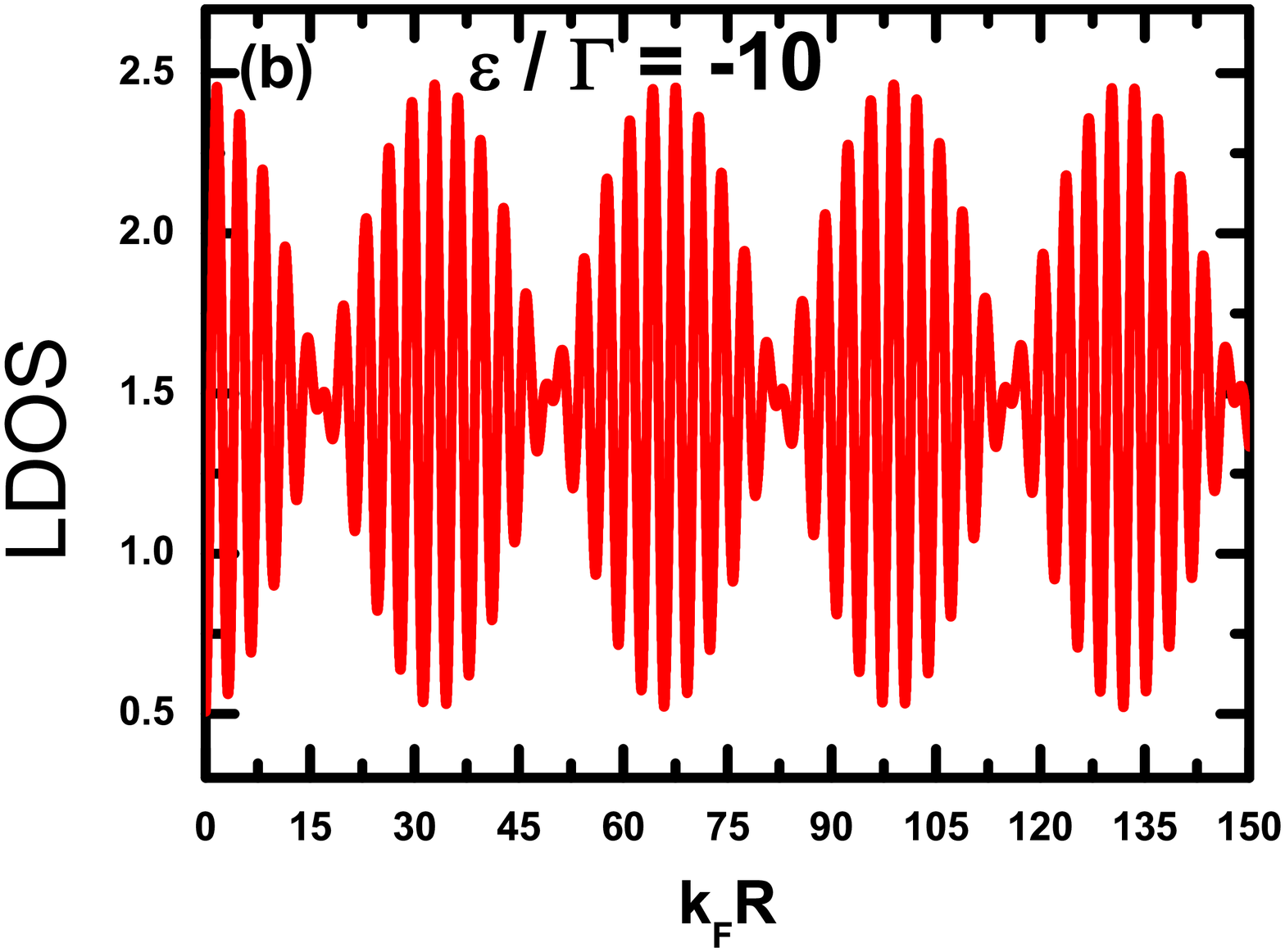}}}
\caption{(Color online) (a) LDOS {[}Eq. (\ref{eq:a_LDOS}){]} at $k_{B}T=0.1\Gamma$
of a quantum wire with $P=0.1$ as a function of $\varepsilon/\Gamma$ for
different values of positions $k_{F}R$. Fano-profiles appear around
$\varepsilon=\varepsilon_{1d}=-10\Gamma$ and $\varepsilon=\varepsilon_{2d}=-4.5\Gamma$.
This pair of antiresonances (black line) can be tuned to resonances
(green line) by moving laterally the STM tip. (b) In opposition to
the metallic surface device, undamped spin-polarized quantum beats in the LDOS at $\varepsilon=\varepsilon_{1d}=-10\Gamma$
manifest.}

\label{Fig4}
\end{figure}

\subsection{Metallic Surface}

Defining $k_{F\downarrow}=k_{F}$, we begin the analysis in the metallic surface
apparatus by dividing our study into regions we call short, intermediate
and long-range limits. The short-range limit presented in Fig. \ref{Fig2}(a)
reveals that the LDOS given by Eq. (\ref{eq:a_LDOS}) as function
of energy exhibits two Fano antiresonances. Each one corresponds
to the discrete levels of the adatom ($j=1$) and the subsurface impurity ($j=2$). The
main feature in this situation is that the Fano profile conserves
its line shape when the dimensionless parameter $k_{F}R$ is changed.
Additionally, this profile is suppressed for increasing distances,
tending to the DOS background of the host.

In Fig. \ref{Fig2}(b) we look at how the LDOS evolves with $k_{F}R$
exactly at the Fano antiresonance $\varepsilon=\varepsilon_{d_{1}}=-10\Gamma$.
At the host site ($k_{F}R=0$), the LDOS presents a depletion. Such a dip in the LDOS is a result of charge screening
around the impurities by conduction electrons, which suppresses
the LDOS of the host. Beyond the adatom position, the LDOS is indeed
dictated by Friedel oscillations, which also lead to a strong decay
in the long-range limit {[}see the Fig. \ref{Fig2}(b){]}. The evanescent
feature of the LDOS is a result of the interplay between the Friedel-like
expression $A_{j\sigma}^{2D}$ and the Fano parameter $q_{j\sigma}^{2D}$.
These quantities are governed by Eqs. (\ref{eq:_soma1_}) and (\ref{eq:Fano_j_2}),
where the former evolves spatially according to the zeroth-order Bessel function
$J_{0}$. Such damping in the LDOS has been already observed
experimentally in a system composed by an Fe host and a Co adatom.\cite{Kawahara}

In Fig. \ref{Fig3}(a) we plot the Fano line shape in the long-range
regime ($k_{F}R>15$). The same dips at $\varepsilon=-10\Gamma$ and
$\varepsilon=-4.5\Gamma$ are observed as in the short-range limit
{[}Fig. \ref{Fig2}(a){]}. However, a contrasting feature is found
between these two limits. While in the short-range case the dips become
suppressed as $k_{F}R$ increases, in the long-range limit the dip
oscillates with $k_{F}R$. This is a result of the oscillatory profile
observed in the LDOS for increasing $k_{F}R$ {[}see Fig. \ref{Fig2}(b){]}.
The oscillations of the dip can be more clearly visualized in Fig.
\ref{Fig3}(b), where we show the LDOS at $\varepsilon=\varepsilon_{d_{1}}=-10\Gamma$.
A peculiar beating is observed in the LDOS due to the slightly different
Fermi wave numbers {[}see Eq. (\ref{eq:kFs}){]}.

\subsection{Quantum Wire}

Figure \ref{Fig4}(a) shows the LDOS plotted against energy for different
$k_{F}R$ values in the short-range limit. For the STM tip at $k_{F}R=0$,
the LDOS shows the two-dip structure already observed in Fig. \ref{Fig2}(a).
In contrast, as $k_{F}R$ increases, the antiresonances change
to resonances, passing through intermediate profiles (asymmetric Fano
line shapes). We emphasize that this behavior in the LDOS was recently observed
in the experiment performed by Pr\"user \textit{et al.} with atoms of Fe and Co
beneath the Cu(100) surface.\cite{STM16}

We observe in Fig. \ref{Fig4}(b) the evolution of the LDOS with $k_{F}R$.
Non-evanescent oscillations occur, modulated by an amplitude beating.
This undamped behavior is encoded by Eqs. (\ref{eq:soma_2}) and
(\ref{eq:fano_1d}), for Friedel-like oscillations ($A_{j\sigma}^{1D}$)
and Fano interference ($q_{j\sigma}^{1D}$), respectively. These quantities
are simple trigonometric functions without damping. This feature
is due to the absence of an extra dimension for the scattering of the electronic wave. On the
other hand, in 2D this propagation is spread in a plane leading to a spatial decay in the LDOS.
Thus the amplitude of the undamped beats is much larger than in the metallic surface device. This means
that in such case, the LDOS signal can be more easily resolved experimentally.

\begin{figure}[h]
 \centerline{\resizebox{3.5in}{!}{ \includegraphics[clip,width=0.55\textwidth]{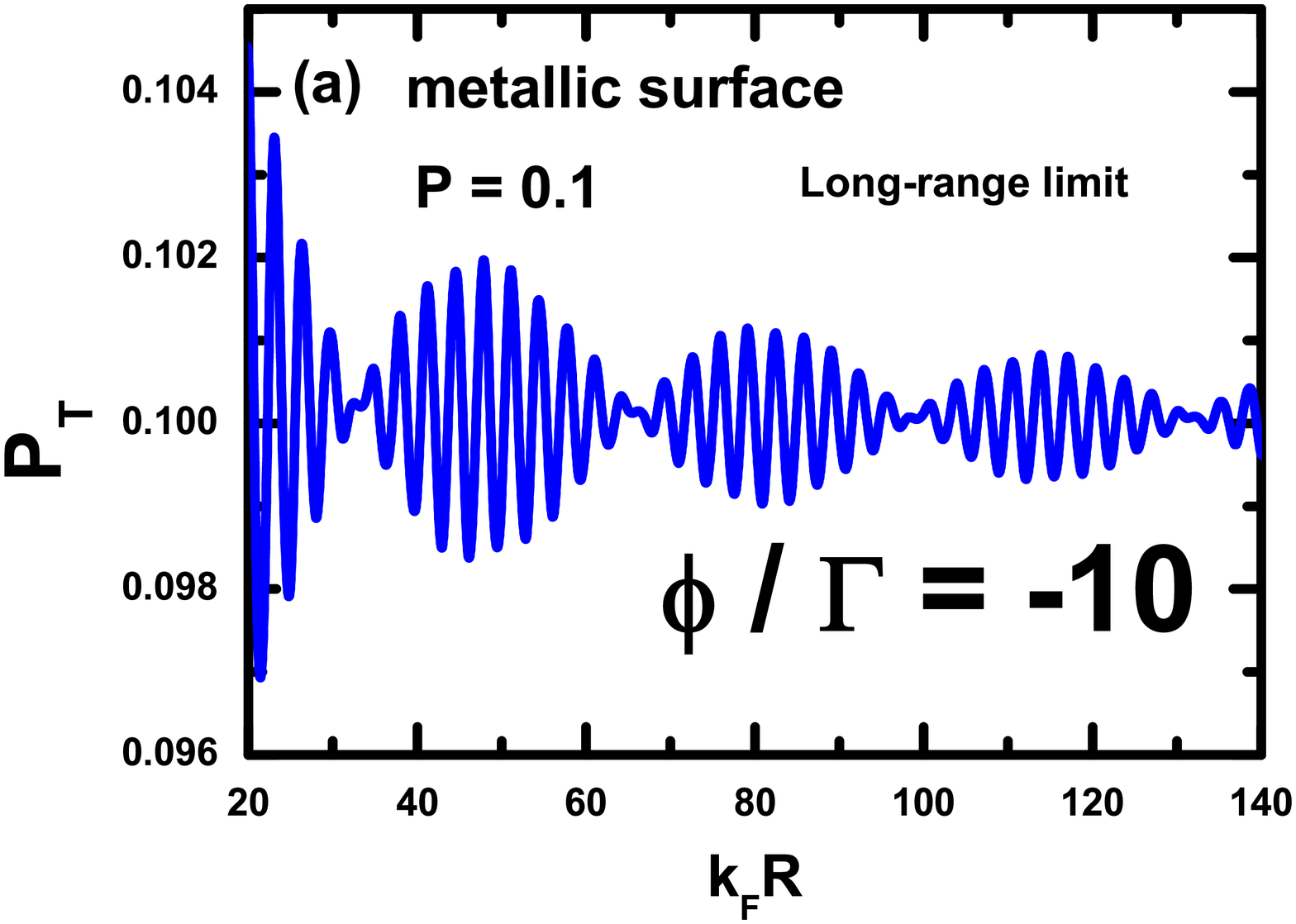}}}
\centerline{\resizebox{3.5in}{!}{ \includegraphics[clip,width=0.55\textwidth]{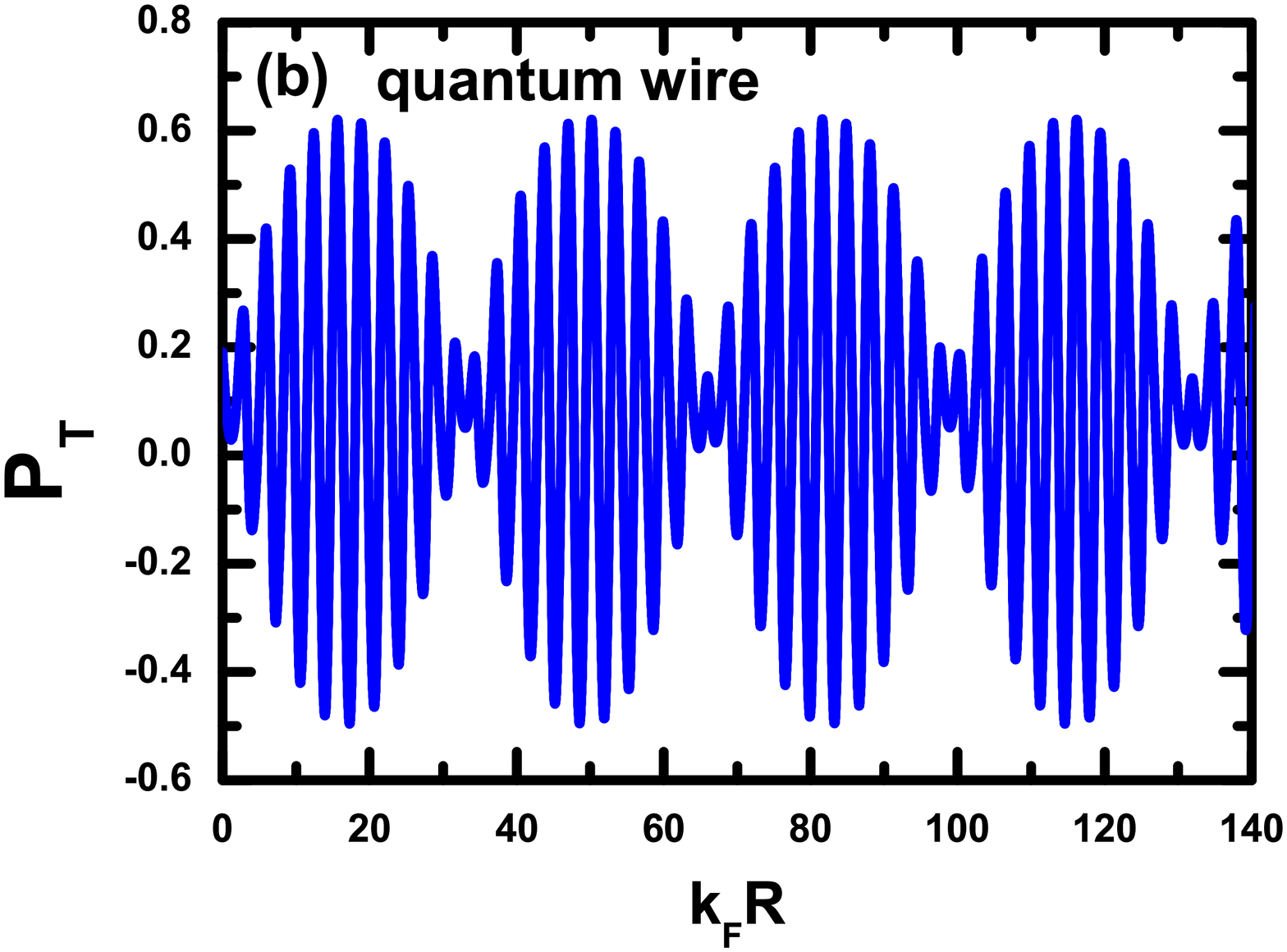}}}
\caption{(Color online) Transport polarization of the ferromagnetic hosts {[}Eq. (\ref{eq:SP_full}){]}
at $k_{B}T=0.1\Gamma$ and applied bias $\phi=\varepsilon_{1d}=-10\Gamma$.
(a) Damped spin-polarized quantum beats appear in the polarization of the metallic surface device. (b)
Undamped type occur in the polarization of the quantum wire. In both situations,
we have a spatially resolved spin-filter, with a polarization that
oscillates around $P=0.1$. In the quantum wire case, this oscillation is more
pronounced.}

\label{Fig5}
\end{figure}

\subsection{Transport Polarization}

Another quantity we investigate is the spin-polarization given by
Eq. (\ref{eq:SP_full}). As the differential conductance of Eq. (\ref{eq:DC})
is proportional to the LDOS {[}Eq. (\ref{eq:a_LDOS}){]}, the ferromagnetic hosts
filter electrons that tunnels into (or out of) the STM tip. This filtering
is dominated by the majority spin component. Thus devices without impurities behave as spin filters with a spatially
uniform polarization that coincides with the value given by Eq. (\ref{eq:SP}).
Here we adopt $P=0.1$. Due to the impurities in the side-coupled geometry and the host dimensionality, this polarization is perturbed
in two different forms. In both the metallic surface and the quantum wire as we can see in
Figs. \ref{Fig5}(a) and \ref{Fig5}(b) with applied bias $\phi=\varepsilon_{1d}=-10\Gamma$,
the polarization oscillates around $P_{T}=P=0.1$. Unlike the metallic surface system,
where small deviations with damping occur, the {}``electron focusing''
effect in the quantum wire leads to undamped and pronounced oscillations. The
polarization in the latter case does not exceed $P_{T}\approx+0.62$
or fall below $P_{T}\approx-0.5$. Therefore the polarized current
through the junction formed by the STM tip and the surface alternates
from spins up ($+0.62$) to down ($-0.5$) depending on the tip position.
Additionally, along this probing direction, the polarization not only can invert the orientation of the majority spin component, but also
becomes zero at some sites, where locally the unbalance of spins is
totally suppressed. As a result, we have a tunneling current without
polarization in specific positions on the sample surface. On
the other hand, as we can see in Fig. \ref{Fig5}(a), the amplitude
of the beats in the metallic surface polarization is extremely suppressed and does
not change its signal $\left(P_{T}>0\right)$. Thus the quantum wire operates
as a spatially resolved spin filter, with a higher efficiency.

\section{Conclusions}

\label{sec5}

In order to investigate a ferromagnetic system with two impurities,
we have calculated, the LDOS and the spin polarization of hosts in two different dimensionalities.
Impurities in the side-coupled geometry, as outlined in Fig. \ref{fig:Pic1},
were taken into account. We analyzed both a metallic surface and a quantum wire described by the two-impurity Anderson model in the picture of a spin-polarized
electron gas with impurities away from the Kondo regime. We presented
a model in which an unperturbed 1D electron host in the presence of localized
states produces undamped behavior in the LDOS Fano profile {[}see
Figs. \ref{Fig4}(a) and \ref{Fig4}(b){]}, similar to that observed
experimentally. \cite{STM16} In contrast, our 2D model revealed a
damped oscillatory behavior {[}Figs. \ref{Fig2}(a), \ref{Fig2}(b),
\ref{Fig3}(a) and \ref{Fig3}(b){]}. We demonstrated that these opposed
features originate from the interplay between the Friedel-like function
and the Fano parameter, which assume different functional forms according
to the host dimensionality. Keeping the energy fixed and tuning the
STM tip position, we verified the emergence of spin-polarized quantum beats in the LDOS
given by Eq. (\ref{eq:a_LDOS}) as well as in the transport polarization
of Eq. (\ref{eq:SP_full}). Such an effect is due to interference
between the slightly different Fermi wave numbers $k_{F\uparrow}$
and $k_{F\downarrow}$ {[}Eq. (\ref{eq:kFs}){]} in the LDOS, achievable
in hosts with low spin polarizations. Therefore the quantum wire setup behaves
as a spatially resolved spin-filter with a high efficiency, as we can
see in Fig. \ref{Fig5}(b). Away from the adatom, this device can
magnify or invert locally the original spin orientation of the host,
also displaying sites where this polarization is completely quenched.
As a possible experimental implementation of this apparatus, we suggest
the systems investigated by Pr\"user. \cite{STM16} Such setups present
the same one-dimensional character as our effective quantum wire model.

\begin{acknowledgments}
This work was supported by the Brazilian agencies CNPq, CAPES,
FAPEMIG and PROPe/UNESP. J. C. Egues also acknowledges PRP/USP within the
Research Support Center Initiative (NAP Q-NANO). \end{acknowledgments}

\end{document}